\PassOptionsToPackage{unicode}{hyperref}
\PassOptionsToPackage{hyphens}{url}
\PassOptionsToPackage{dvipsnames,svgnames,x11names}{xcolor}
\documentclass[12pt]{article}

\usepackage{amsmath,amssymb}
\usepackage{iftex}
\usepackage{bm}
\usepackage{algorithm}
\usepackage[noend]{algpseudocode}
\usepackage{multirow}      
\usepackage{dcolumn}       
\usepackage{makecell}      
\usepackage{rotating}      
\usepackage{tablefootnote}
\usepackage{subfigure}

\ifPDFTeX
  \usepackage[T1]{fontenc}
  \usepackage[utf8]{inputenc}
  \usepackage{textcomp} 
\else 
  \usepackage{unicode-math}
  \defaultfontfeatures{Scale=MatchLowercase}
  \defaultfontfeatures[\rmfamily]{Ligatures=TeX,Scale=1}
  \renewcommand{\bm}[1]{\symbf{#1}}
  \renewcommand{\boldsymbol}[1]{\symbf{#1}}
\fi
\usepackage{lmodern}
\ifPDFTeX\else  
\fi
\IfFileExists{upquote.sty}{\usepackage{upquote}}{}
\IfFileExists{microtype.sty}{
  \usepackage[]{microtype}
  \UseMicrotypeSet[protrusion]{basicmath} 
}{}
\makeatletter
\@ifundefined{KOMAClassName}{
  \IfFileExists{parskip.sty}{%
    \usepackage{parskip}
  }{
    \setlength{\parindent}{0pt}
    \setlength{\parskip}{6pt plus 2pt minus 1pt}}
}{
  \KOMAoptions{parskip=half}}
\makeatother
\usepackage[table]{xcolor}
\makeatletter
\ifx\paragraph\undefined\else
  \let\oldparagraph\paragraph
  \renewcommand{\paragraph}{
    \@ifstar
      \xxxParagraphStar
      \xxxParagraphNoStar
  }
  \newcommand{\xxxParagraphStar}[1]{\oldparagraph*{#1}\mbox{}}
  \newcommand{\xxxParagraphNoStar}[1]{\oldparagraph{#1}\mbox{}}
\fi
\ifx\subparagraph\undefined\else
  \let\oldsubparagraph\subparagraph
  \renewcommand{\subparagraph}{
    \@ifstar
      \xxxSubParagraphStar
      \xxxSubParagraphNoStar
  }
  \newcommand{\xxxSubParagraphStar}[1]{\oldsubparagraph*{#1}\mbox{}}
  \newcommand{\xxxSubParagraphNoStar}[1]{\oldsubparagraph{#1}\mbox{}}
\fi
\makeatother

\usepackage{longtable,booktabs,array}
\usepackage{calc} 
\usepackage{etoolbox}
\makeatletter
\patchcmd\longtable{\par}{\if@noskipsec\mbox{}\fi\par}{}{}
\makeatother
\IfFileExists{footnotehyper.sty}{\usepackage{footnotehyper}}{\usepackage{footnote}}
\makesavenoteenv{longtable}
\usepackage{graphicx}
\makeatletter
\def\maxwidth{\ifdim\Gin@nat@width>\linewidth\linewidth\else\Gin@nat@width\fi}
\def\maxheight{\ifdim\Gin@nat@height>\textheight\textheight\else\Gin@nat@height\fi}
\makeatother
\setkeys{Gin}{width=\maxwidth,height=\maxheight,keepaspectratio}
\makeatletter
\def\fps@figure{htbp}
\makeatother

\makeatletter
\@ifpackageloaded{caption}{}{\usepackage{caption}}
\AtBeginDocument{%
\ifdefined\contentsname
  \renewcommand*\contentsname{Table of contents}
\else
  \newcommand\contentsname{Table of contents}
\fi
\ifdefined\listfigurename
  \renewcommand*\listfigurename{List of Figures}
\else
  \newcommand\listfigurename{List of Figures}
\fi
\ifdefined\listtablename
  \renewcommand*\listtablename{List of Tables}
\else
  \newcommand\listtablename{List of Tables}
\fi
\ifdefined\figurename
  \renewcommand*\figurename{Figure}
\else
  \newcommand\figurename{Figure}
\fi
\ifdefined\tablename
  \renewcommand*\tablename{Table}
\else
  \newcommand\tablename{Table}
\fi
}
\@ifpackageloaded{float}{}{\usepackage{float}}
\floatstyle{ruled}
\@ifundefined{c@chapter}{\newfloat{codelisting}{h}{lop}}{\newfloat{codelisting}{h}{lop}[chapter]}
\floatname{codelisting}{Listing}

\makeatother
\makeatletter
\@ifpackageloaded{caption}{}{\usepackage{caption}}
\@ifpackageloaded{subcaption}{}{\usepackage{subcaption}}
\makeatother

\ifLuaTeX
  \usepackage{selnolig}  
\fi
\usepackage[]{natbib}
\usepackage{bookmark}

\IfFileExists{xurl.sty}{\usepackage{xurl}}{} 
\hypersetup{
  pdftitle={Directed hypergraphs: tensor representation, embedding modeling, and community detection},
  pdfauthor={Guojun Zhu; Haofei Wang; Mingyang Ren},
  pdfkeywords={Directed graph; Hypergraph; Latent space model; Representation learning},
  colorlinks=true,
  linkcolor={blue},
  filecolor={Maroon},
  citecolor={Blue},
  urlcolor={Blue},
  pdfcreator={LaTeX via pandoc}}

\definecolor{LightPink}{RGB}{255,228,225}
\DeclareMathOperator*{\argmin}{arg\,min}
\setcitestyle{authoryear,open={(},close={)}}
\hypersetup{
    colorlinks=true,
    linkcolor=blue,
    anchorcolor=blue,
    urlcolor=blue,
    citecolor=blue
}

\counterwithout{Lem}{section}
\newtheorem{Con}{Condition}
\counterwithout{Con}{section}
\newtheorem{Mth}{Theorem}[section]
\counterwithout{Mth}{section}
\newtheorem{Rem}{Remark}[section]
\counterwithout{Rem}{section}

\counterwithout{Def}{section}
\newtheorem{Cor}{Corollary}[section]
\counterwithout{Cor}{section}

\counterwithout{Pro}{section}
\numberwithin{equation}{section}

\def\argmin{\mathop{\rm argmin}}

\def\pr{\mathop{\rm pr}}

\newcommand{\anon}{1}

\graphicspath{{figures/}}

\begin{document}

\def\spacingset#1{\renewcommand{\baselinestretch}%
{#1}\small\normalsize} \spacingset{1}

\if1\anon
{
  \title{\bf Directed hypergraphs: tensor representation, embedding modeling, and community detection}
  
  \author{Guojun Zhu$^{1,2}$, Haofei Wang$^{3}$ and Mingyang Ren$^{3,\dagger}$\\
    \small $^1$ School of Mathematical Sciences, University of Chinese Academy of Sciences \\
    \small $^2$ Department of Statistics \& Data Science, National University of Singapore \\
    \small $^3$ School of Mathematical Sciences, Shanghai Jiao Tong University \\
    \small $^\dagger$ Corresponding author: mingyangren@sjtu.edu.cn
    }
    \date{}
  \maketitle
} \fi

\if0\anon
{
  \bigskip
  \bigskip
  \bigskip
  \begin{center}
    {\LARGE\bf Directed hypergraphs: tensor representation, embedding modeling, and community detection}
\end{center}
  \medskip
} \fi

\bigskip
\begin{abstract}
Directed hypergraphs, in which each hyperedge connects a set of tail nodes to a set of head nodes, are the natural representation for many-to-many interactions in citation networks, metabolic pathways, and group communications. Despite their broad applicability, community detection for directed hypergraphs remains statistically underdeveloped: most existing methods either collapse hyperedges into pairwise edges, treat the two sides as interchangeable, or require the number of communities to be known. We propose a statistical framework for simultaneous embedding and community detection in nonuniform directed hypergraphs. The method allows each node to have distinct roles as a source and as a target, accommodates varying hyperedge sizes, and automatically determines the unknown numbers of communities on both sides, which need not be equal, through a fusion-penalized likelihood. We establish consistency of the estimated numbers of communities, exact recovery of both membership structures, and consistent probability tensor estimation. We further show that exact recovery is attainable without knowing the numbers of communities by a procedure that does not rely on any particular local solution, and we complement the upper bounds with a minimax lower bound for probability tensor estimation and a necessary condition for exact recovery. To ground the methodology, we construct the Cross-Statistics Citation Network (CSCN), a new large-scale corpus linking Statistics with Biology, Computer Science, and Economics. Applying our method to CSCN reveals source and target roles of author communities in cross-field knowledge flow, and a pairwise representation of the same citations partly merges these roles.
\end{abstract}

\noindent%
\vfill

\newpage
\spacingset{1.8} 

\section{Introduction}

Many modern datasets record relations among interacting units rather than independent observations. Networks provide a natural framework for such data, and a central inferential task is community detection: grouping nodes with similar connection patterns to reveal functional organization or structural heterogeneity that is difficult to discern from individual connections alone \citep{newman2003structure,goldenberg2010survey}. 
Most statistical approaches formulate community detection on ordinary graphs, where each observed relation is represented by an edge between two nodes. Undirected graphs model symmetric relations such as collaboration, while directed graphs retain an ordered tail and head and are suitable for phenomena like paper citation. Substantial progress has been made under stochastic block models \citep{karrer2011stochastic,wang2023fast}, degree-corrected variants \citep{ji2022cocitation,jiang2024optimal}, and directed models that allow asymmetric tail and head connection patterns \citep{zhang2022directed,su2024spectral}. Recent methods also estimate the number of communities rather than requiring it as input \citep{jin2023optimal,ren2023consistent,zhang2025network}. 

An ordinary graph, however, remains intrinsically pairwise: every edge involves exactly two nodes. When a single event jointly involves multiple units, representing it by pairwise edges discards which units co-participated in the same event, removes the boundary between distinct group interactions, and gives disproportionate weight to events involving larger groups. Motif-based methods can incorporate higher-order information into community recovery \citep{yu2024higherorder}, but their observational units remain dyadic. Directionality alone does not resolve this limitation, as it preserves ordering between two nodes but not the many-to-many structure of the original event.

These considerations motivate moving from graphs to hypergraphs, where one hyperedge may contain any number of nodes and thus retains both the composition and size of a group interaction, as in multi-author collaborations, group communications, and protein complexes \citep{benson2016higher,battiston2020networks}. Statistical community detection for hypergraphs has been developed through spectral partitioning \citep{ghoshdastidar2017consistency}, tensor-based approaches \citep{Ke2019Community}, latent-space models \citep{lyu2023latent,yu2025modeling,chen2026modeling}, and embedding methods \citep{zhen2023community}. Recent probabilistic block models estimate hard or overlapping memberships directly from observed hyperedges \citep{chodrow2021generative,contisciani2022inference,ruggeri2023community}, and complementary theory studies testing and exact recovery under hypergraph or tensor block models \citep{yuan2022testing,han2022exact}. This literature shows that retaining higher-order interactions can recover organization that graph projections obscure. Information-theoretic limits are also available for hypergraph stochastic block models. Sharp thresholds for exact recovery are known for uniform \citep{zhang2023exact} and nonuniform \citep{dumitriu2026optimal} hypergraphs, extending those for graphs \citep{abbe2016exact}, and minimax misclassification rates are known for uniform hypergraphs \citep{chien2019minimax}. 

These works, however, focus on undirected hypergraphs, where nodes within a hyperedge have exchangeable roles. Such a formulation cannot record that one set of nodes acts on, supplies, or precedes another set. Many higher-order relations have exactly this ordered two-set structure. A metabolic reaction transforms a set of reactants into a set of products. A paper citation links the authors of the cited paper (the tail set) to the authors of the citing paper (the head set). Throughout this article, we adopt this directed hyperedge representation, which retains both the citation direction and the paper-level author sets. An undirected hypergraph removes the tail-head ordering. A projected directed graph replaces one many-to-many citation by all author-pair edges, erases the event boundary, and overweights citations involving large teams.

Statistical methods for such ordered many-to-many relations remain scarce. Directed hypergraphs have a well-established combinatorial formulation \citep{gallo1993directed}, and recent work has developed reciprocity measures and generators \citep{kim2023reciprocity} as well as motif-based structural summaries \citep{lotito2026microscale}. Ordinary directed-network models can also distinguish tail and head connection patterns \citep{zhang2022directed,zhang2025network}. However, these developments do not yet provide a community-detection framework that (i) models many-to-many hyperedges without pairwise projection; (ii) permits distinct tail and head community structures with unknown, possibly unequal numbers; (iii) estimates hyperedge probabilities; and (iv) provides statistical recovery guarantees in sparse regimes. 

This methodological gap is accompanied by a data limitation. Existing scholarly network analyses are often based on authors, papers, or pairwise co-citation and coauthorship relations within a single discipline or a restricted set of venues \citep{ji2022cocitation}. There is still no large-scale, statistics-centered corpus that records how research in Statistics cites, and is cited by, work from multiple neighboring fields. The absence of such data makes it difficult to study cross-field knowledge flow as a directional group-level phenomenon.

To address both limitations, we construct the {\bf Cross-Statistics Citation Network (CSCN)} dataset, which links Statistics with Biology, Computer Science, and Economics (Section \ref{sec:real-data}). 
\begin{figure}[h]
    \centering
    \includegraphics[width=0.9\textwidth]{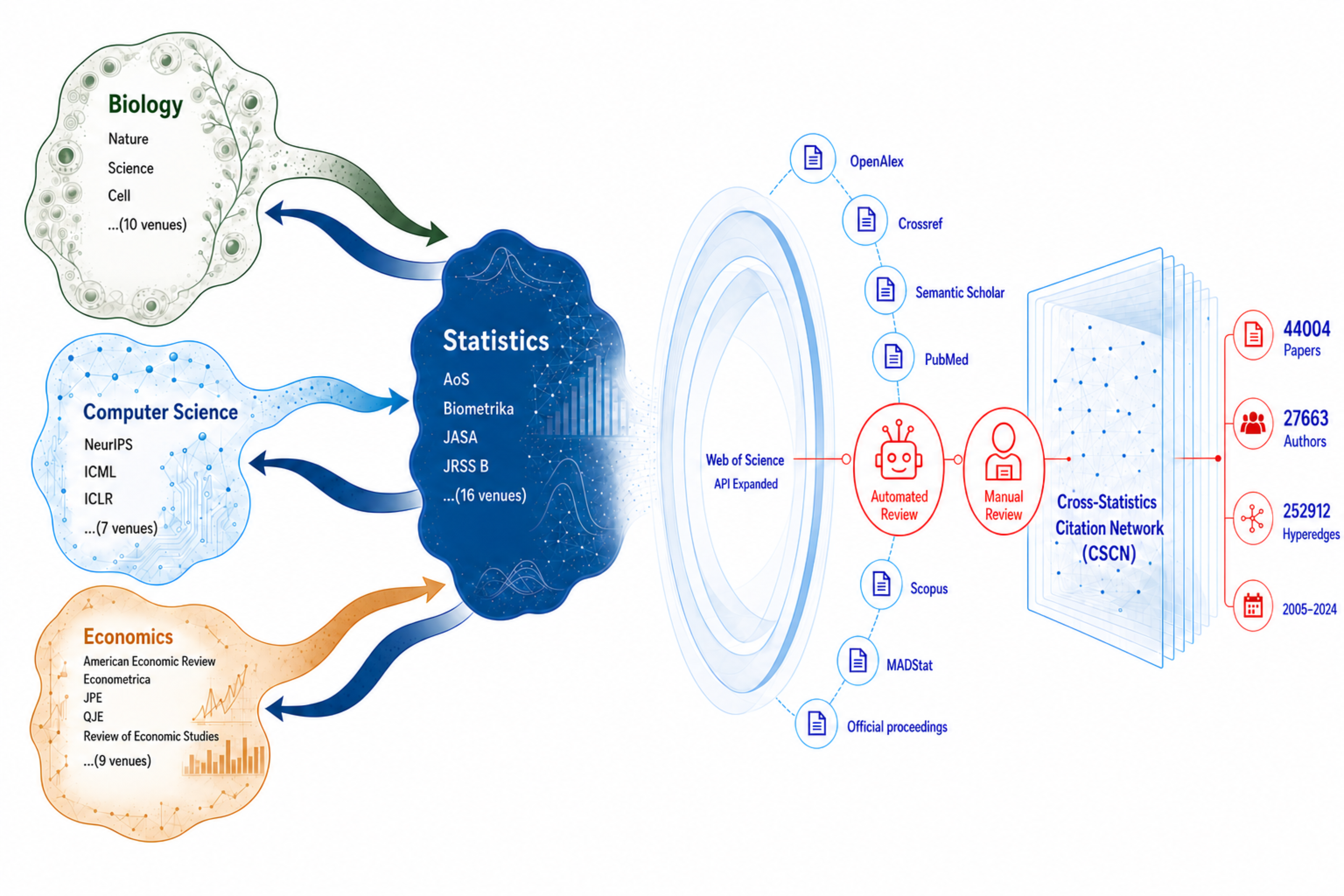}
    \caption{Overview of the CSCN dataset.}
    \label{fig:cscn-schematic}
\end{figure}
CSCN preserves each paper citation as a relation from the cited-paper author set to the citing-paper author set, so that knowledge supply and uptake can be studied as distinct group roles and an author may occupy different communities on the two sides. It therefore provides a new resource for studying cross-field knowledge flow and an empirical setting that calls for directed-hypergraph community inference. Figure~\ref{fig:cscn-schematic} summarizes venue scope, data sources, and contents of CSCN.

To address the methodological gap, we develop a statistical embedding and community-detection framework for nonuniform directed hypergraphs. Each directed hyperedge is an ordered pair consisting of a tail set and a head set, and each real node has separate latent embeddings for its tail and head roles. The two embeddings allow a node to belong to different communities on the two sides, and the numbers of tail and head communities need not agree. To accommodate varying hyperedge sizes, we pad short tails and heads with distinct null nodes. This augmentation converts the observed hypergraph into a fixed-order adjacency tensor that is symmetric within each side but retains the ordering between the two sides. We then impose a low-rank model with separate tail and head factors. Our procedure, {\bf FOLD} ({\bf F}usion-based tw{\bf o}-sided {\bf L}atent embedding for {\bf D}irected hypergraphs), links these factors to node-specific centers and fuses nearby centers on each side. The distinct fitted centers determine the memberships and the two numbers of communities without requiring either number in advance. We compute the estimator by alternating gradient updates for the embeddings with ADMM updates for the fused centers, and select the penalty parameters by a data-driven criterion. The same formulation supports both probability estimation and community recovery from a single fit.

The proposed framework contributes to the literature in four ways. First, we develop a tensor-based representation for nonuniform directed hypergraphs via null-node augmentation, and a low-rank embedding model that separately parametrizes tail-side and head-side interactions. This representation preserves the ordered many-to-many structure of directed hyperedges and naturally accommodates varying hyperedge cardinalities without resorting to pairwise projection or ad hoc symmetrization. Second, we introduce a fusion-penalized estimation procedure that simultaneously recovers the tail and head community structures and determines their numbers automatically. By coupling each node's embedding to a latent center and fusing centers within each side, FOLD does not require the number of communities on either side to be specified, a requirement that limits most existing hypergraph and directed-network methods. The alternating optimization scheme combines the scalability of gradient-based updates with the structured shrinkage of ADMM. Third, we establish that the estimated numbers of tail and head communities are consistent, that both membership partitions are recovered exactly with probability tending to one, and that the underlying probability tensor is estimated consistently. These results extend statistical recovery guarantees from ordinary directed graphs and undirected hypergraphs to nonuniform many-to-many interactions with unknown and potentially unequal numbers of communities. We further show that both partitions and their numbers of communities are recovered exactly by a procedure that does not rely on any particular local minimizer of the nonconvex objective. We also prove a minimax lower bound, which shows that the estimation rates are optimal up to a logarithmic factor, and a necessary condition for exact recovery. Fourth, we construct the Cross-Statistics Citation Network (CSCN), a new large-scale corpus that connects Statistics with Biology, Computer Science, and Economics. Applying FOLD to CSCN reveals asymmetric knowledge roles that an undirected partition cannot represent and a pairwise projection of the same citations partly merges.

\section{Methodology}

Directed hypergraphs combine higher-order interactions with asymmetric roles. A statistical representation must retain the ordering from tail to head, remain invariant to permutations within each side, and accommodate hyperedges whose tail and head cardinalities vary.

\subsection{Null-node augmentation and tensor representation}

\begin{figure}[H]
    \centering
    \subfigure[Nonuniform directed hypergraph.]{
    \includegraphics[height=0.245\textheight]{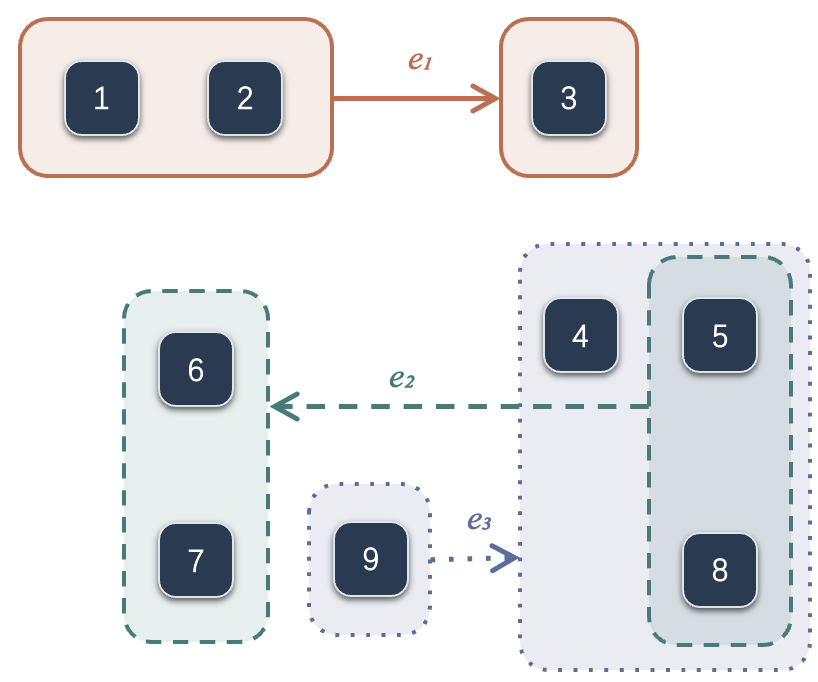}}
    \hspace{0.03\textwidth}
    \subfigure[$(2,3)$-uniform augmentation.]{
    \includegraphics[height=0.245\textheight]{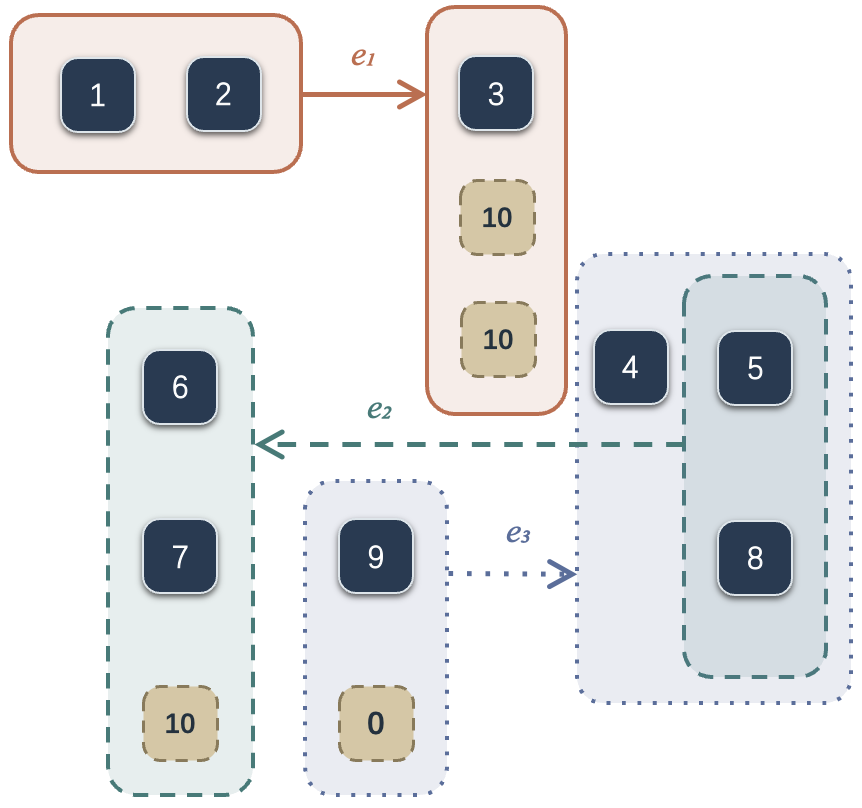}}
    \caption{Null-node augmentation of a nonuniform directed hypergraph. Nodes $0$ and $10$ are the tail-null and head-null nodes, respectively.}
    \label{fig:directed-hypergraph-augmentation}
\end{figure}

Let the set of real nodes be $V=[n]=\{1,\ldots,n\}$. A directed hyperedge is an ordered pair $e=(T_e, H_e)$, where $T_e\subseteq [n]$ is a tail set, $H_e\subseteq [n]$ is a head set, $T_e\neq \varnothing$, $H_e\neq \varnothing$, and $T_e\cap H_e=\varnothing$. Write $\mathcal{H}=(V,E)$ for a directed hypergraph with edge set $E$. As illustrated in Figure~\ref{fig:directed-hypergraph-augmentation}(a), we allow the edge cardinalities to vary, but assume that there exist two fixed integers $p\geqslant 1$ and $q\geqslant 1$ such that $|T_e|\leqslant p$ and $|H_e|\leqslant q$ for every $e=(T_e,H_e)\in E$, where $p$ and $q$ are the maximal tail and head sizes, respectively. 

This creates two difficulties for statistical modeling.  First, hyperedges with different tail and head cardinalities cannot be directly collected into one fixed-order adjacency tensor.  Second, directionality means that permutation invariance holds only within the tail set and within the head set, but not across the two sides.  We therefore begin with a null-node augmentation step, in the same spirit as the augmentation used for undirected nonuniform hypergraphs \citep{zhen2023community}, but adapted here to preserve the two directional roles.
As illustrated in Figure~\ref{fig:directed-hypergraph-augmentation}(b), we introduce two null nodes: a tail-null node indexed by $0$ and a head-null node indexed by $n+1$. The augmented node set is $\bar V=\{0\}\cup [n]\cup \{n+1\}$, so that the total number of nodes becomes $n+2$. For a directed hyperedge $e=(T_e,H_e)$ with $|T_e|=a\leqslant p$ and $|H_e|=b\leqslant q$, define its augmented version $e^+=(T_e^+,H_e^+)$ by:
\begin{equation}\notag 
    T^+_e=\{\!\{T_e,\underbrace{0,\ldots,0}_{p-a\text{ times}}\}\!\},\qquad H^+_e=\{\!\{H_e,\underbrace{n+1,\ldots,n+1}_{q-b\text{ times}}\}\!\}. 
\end{equation}
Then $T^+_e$ is a multiset of cardinality $p$ and $H^+_e$ is a multiset of cardinality $q$. Only the tail-null node $0$ is allowed to repeat in the tail set, and only the head-null node $n+1$ is allowed to repeat in the head set. In this way, each original directed hyperedge is converted into a $(p,q)$-uniform directed augmented hyperedge.

The natural combinatorial object after augmentation is an ordered pair of multisets rather than a single multiset. More precisely, an augmented directed hyperedge is represented by $(T^+_e,H^+_e)$ with $|T^+_e|=p$ and $|H^+_e|=q$. The ordering between the two sets matters, while the ordering inside each set does not. This is why the adjacency tensor below is only partially symmetric: it is symmetric within the first $p$ modes and within the last $q$ modes, but not across the two sets.

For a tuple $\omega=(i_1,\ldots,i_p; j_1,\ldots,j_q)\in \bar V^{\,p+q}$, call $\omega$ legal if the following conditions hold. The indices $i_1,\ldots,i_p$ belong to $\{0,1,\ldots,n\}$, and the only index allowed to repeat in the tail set is $0$. Moreover, not all $i_1,\ldots,i_p$ are equal to $0$. The indices $j_1,\ldots,j_q$ belong to $\{1,\ldots,n,n+1\}$, and the only index allowed to repeat in the head set is $n+1$. Moreover, not all $j_1,\ldots,j_q$ are equal to $n+1$. No real node is allowed to appear in both sets simultaneously. In particular, the tail-null node never appears in the head set, and the head-null node never appears in the tail set. Define the directed augmented Kronecker delta by $\delta_{\omega}=\mathbb{I}(\omega \text{~is legal})$, where $\mathbb{I}(\cdot)$ is the indicator function. Define also the ordered version by $\delta^{\mathrm{ord}}_{\omega}=\delta_{\omega}\mathbb{I}(i_1\leqslant \cdots \leqslant i_p,\ j_1\leqslant \cdots \leqslant j_q)$. The role of $\delta_{\omega}$ is to restrict attention to tuples that can truly arise from augmented directed hyperedges. The adjacency tensor is an order-$(p+q)$ binary tensor $\bm{A}=\bigl(a_{i_1\cdots i_p\, j_1\cdots j_q}\bigr)_{i_1,\cdots ,i_p \in \bar V \setminus \{n+1\}, j_1,\cdots, j_q \in \bar V \setminus \{0\} } \in \{0 , 1\}^{(n+1) \times \cdots \times(n+1)}$, whose entry is defined by $a_{i_1\cdots i_p\, j_1\cdots j_q} = \mathbb{I}((\{\!\{i_1,\ldots,i_p\}\!\},\{\!\{j_1,\ldots,j_q\}\!\}\bigr)\in E^+)$ where
 $E^+=\{e^+:e\in E\}$ is the set of augmented directed hyperedges. A necessary condition for $a_{i_1\cdots i_p\, j_1\cdots j_q}=1$ is $\delta_{\omega}=1$. By construction, $a_{i_1\cdots i_p\, j_1\cdots j_q}=a_{i_{\pi(1)}\cdots i_{\pi(p)}\, j_{\tau(1)}\cdots j_{\tau(q)}}$ for every permutation $\pi\in S_p$ and $\tau\in S_q$, but in general $a_{i_1\cdots i_p\, j_1\cdots j_q}\neq a_{j_1\cdots j_q\, i_1\cdots i_p}$. Hence $\bm{A}$ is symmetric within the tail and within the head but not across the two sides.

The correct notion of equivalence is therefore the orbit generated by within-set permutations. For two tuples $\omega=(i_1,\ldots,i_p; j_1,\ldots,j_q)$ and $\omega'=(i_1',\ldots,i_p'; j_1',\ldots,j_q')$, write $\omega\sim \omega'$ if there exist permutations $\pi\in S_p$ and $\tau\in S_q$ such that $i'_{\ell}=i_{\pi(\ell)}$ and $j'_m=j_{\tau(m)}$ for all $\ell=1,\ldots,p$ and $m=1,\ldots,q$. Denote the corresponding equivalence class by $[\omega]$. Each legal equivalence class $[\omega]$ uniquely determines one augmented directed hyperedge $\bigl(\{\!\{i_1,\ldots,i_p\}\!\},\{\!\{j_1,\ldots,j_q\}\!\}\bigr)$, and each augmented directed hyperedge determines exactly one such equivalence class. In practice, it is enough to work on the canonical representative set $\Omega_{p,q}^{\mathrm{ord}}=\Bigl\{(i_1,\ldots,i_p; j_1,\ldots,j_q)\in \bar V^{\,p+q}:\delta^{\mathrm{ord}}_{\omega}=1\Bigr\}$. The set $\Omega_{p,q}^{\mathrm{ord}}$ contains exactly one representative for each admissible augmented directed hyperedge.

\subsection{Embedding modeling}

We now specify a probabilistic model for the canonical representatives in $\Omega_{p,q}^{\mathrm{ord}}$. For each legal tuple $\omega=(i_1,\ldots,i_p; j_1,\ldots,j_q)\in\Omega_{p,q}^{\mathrm{ord}}$, let $a_\omega$ denote the corresponding adjacency entry. Conditional on the model parameters, we assume that $a_\omega$ follows a Bernoulli distribution with hyperedge probability $\varrho_\omega$, namely $\pr(a_\omega=1)=\varrho_{\omega}$, and the entries $\{a_\omega:\omega\in\Omega_{p,q}^{\mathrm{ord}}\}$ are modeled independently. To accommodate sparse directed hypergraphs, we use a modified-logit parametrization:
\begin{equation}\notag 
    \theta_{\omega}=\log \frac{\varrho_{\omega}}{s_n-\varrho_{\omega}},\qquad \varrho_{\omega}=\frac{s_n}{1+\exp(-\theta_{\omega})},
\end{equation}
where $s_n\in(0,1]$ is a sparsity factor. Thus, $\theta_\omega$ is the transformed interaction parameter of the augmented hyperedge represented by $\omega$, and $\varrho_\omega$ is its probability.

It remains to impose structure on the collection of transformed interaction parameters $\{\theta_\omega:\omega\in\Omega_{p,q}^{\mathrm{ord}}\}$. For each tail-node index $i\in\{0\}\cup[n]$, introduce a tail embedding $\bm\beta_i\in\mathbb R^r$. For each head-node index $j\in[n+1]$, introduce a head embedding $\bm\zeta_j\in\mathbb R^r$. 
Let $\bm{\beta}=\bigl(\bm{\beta}_0,\bm{\beta}_1,\ldots,\bm{\beta}_{n}\bigr)^\top \in \mathbb{R}^{(n+1)\times r}$ and $\bm{\zeta}=\bigl(\bm{\zeta}_1,\bm{\zeta}_2,\ldots,\bm{\zeta}_{n+1}\bigr)^\top \in \mathbb{R}^{(n+1)\times r}$. We fix the null embeddings at $\bm{\beta}_0=r^{-1/2}\bm{1}_r$ and $\bm{\zeta}_{n+1}=r^{-1/2}\bm{1}_r$, so that short tails and short heads can be represented within the same order-$(p+q)$ model without additional parameters.

Let $\mathcal{I}\in \mathbb{R}^{r\times\cdots\times r}$ be the $(p+q)$th-order identity tensor of dimension $r$, that is, $\mathcal{I}_{k_1\cdots k_{p+q}}=\mathbb{I}( k_1=\cdots=k_{p+q})$. We impose the following low-rank structure on the transformed interaction tensor:
\begin{equation}\notag 
    \bm{\Theta}=\mathcal{I}\times_1 \bm{\beta}\times_2 \cdots\times_p \bm{\beta}\times_{p+1} \bm{\zeta}\times_{p+2} \cdots\times_{p+q} \bm{\zeta}. 
\end{equation}
Let $\bm{\beta}_{\cdot k}$ and $\bm{\zeta}_{\cdot k}$ denote the $k$th columns of $\bm{\beta}$ and $\bm{\zeta}$, and let $\circ$ denote the outer product. The structure above is the CP decomposition
\begin{equation}\notag
    \bm{\Theta}=\sum_{k=1}^{r}\underbrace{\bm{\beta}_{\cdot k}\circ\cdots\circ\bm{\beta}_{\cdot k}}_{p}\circ\underbrace{\bm{\zeta}_{\cdot k}\circ\cdots\circ\bm{\zeta}_{\cdot k}}_{q},
\end{equation}
so that $\bm{\Theta}$ has CP rank at most $r$. Entrywise, for any $\omega=(i_1,\ldots,i_p; j_1,\ldots,j_q)\in\Omega_{p,q}^{\mathrm{ord}}$,
\begin{equation}\notag 
    \theta_{\omega}=\sum_{k=1}^r \beta_{i_1k}\cdots \beta_{i_pk}\,\zeta_{j_1k}\cdots \zeta_{j_qk}. 
\end{equation}
The first $p$ factors encode tail-side participation through $\bm{\beta}$, whereas the last $q$ factors encode head-side participation through $\bm{\zeta}$. This construction can be viewed as a directed analogue of the undirected rank-$r$ hypergraph embedding model, with separate embeddings for tail and head roles. Giving each node a tail copy and a head copy does not reduce the model to the undirected model of \citet{zhen2023community} on $2n$ nodes, because a node cannot appear in both the tail and the head of one hyperedge and every hyperedge has a nonempty tail and a nonempty head. The likelihood below is accordingly defined only on the legal tuples in $\Omega_{p,q}^{\mathrm{ord}}$. Let $u_\omega$ denote the number of null positions of $\omega$. Since the null embeddings are $r^{-1/2}\bm{1}_r$,
\begin{equation}\notag
    \theta_{\omega}=r^{-u_\omega/2}\sum_{k=1}^r\prod_{u:\,i_u\in[n]}\beta_{i_uk}\prod_{v:\,j_v\in[n]}\zeta_{j_vk},
\end{equation}
that is, a hyperedge of any size is modeled by the same CP form over its real nodes, up to a factor that depends only on its size. The identity tensor $\mathcal{I}$ can be replaced by a diagonal tensor with diagonal entries $d_1,\ldots,d_r\in\{-1,1\}$, which gives $\theta_{\omega}=\sum_{k=1}^r d_k\beta_{i_1k}\cdots \beta_{i_pk}\,\zeta_{j_1k}\cdots \zeta_{j_qk}$ and lets each rank-one component enter with either sign. The theoretical results in Section \ref{sec:theory} hold for this model with the same proofs, provided that the same diagonal tensor is used in the definition of $\bm{W}^{*}$. We work with $\mathcal{I}$ throughout. Since each augmented directed hyperedge corresponds to one canonical representative in $\Omega_{p,q}^{\mathrm{ord}}$, the negative log-likelihood can be written as:
\begin{equation}\notag 
    L(\bm{\beta},\bm{\zeta};\bm{A})=-\frac{1}{\phi(n, p, q)}\sum_{\omega\in \Omega_{p,q}^{\mathrm{ord}}}\ell (\theta_{\omega}; a_\omega),
\end{equation}
where $\phi(n, p, q) = |\Omega_{p,q}^{\mathrm{ord}}|$ and $\ell (\theta_{\omega}; a_\omega) = a_\omega\log \varrho_{\omega}+\bigl(1-a_\omega\bigr)\log \bigl(1-\varrho_{\omega}\bigr)$, with $\varrho_\omega=s_n\{1+\exp(-\theta_\omega)\}^{-1}$.

The embedding model also connects probabilistic modeling with community detection. Under the low-rank parametrization above, the hyperedge probability is determined by the interaction between the tail embeddings and the head embeddings. Hence, nodes with nearby tail embeddings tend to have similar tail-side interaction profiles, and nodes with nearby head embeddings tend to have similar head-side interaction profiles. In this sense, clustering in the latent embedding spaces corresponds to grouping nodes with similar roles in the directed hypergraph. We therefore use the latent-space structure not only to estimate hyperedge probabilities, but also to obtain a more interpretable community-level summary of the directed hypergraph.
    
A tail community consists of nodes with similar tail-side interaction patterns, and a head community is defined analogously through head-side interaction patterns. Let $K_{\mathrm{ta}}$ and $K_{\mathrm{he}}$ denote the numbers of tail and head communities, respectively. Prior work on community detection typically requires $K_{\mathrm{ta}}$ and $K_{\mathrm{he}}$ to be specified in advance. To avoid fixing these two quantities beforehand, we introduce node-specific tail and head centers together with fusion penalties, so that the numbers of distinct fitted centers determine the numbers of communities automatically.

Specifically, let $\bm{B}=\bigl(\bm{b}_1,\ldots,\bm{b}_n\bigr)^\top \in \mathbb{R}^{n\times r}$ and $\bm{Z}=\bigl(\bm{z}_1,\ldots \bm{z}_n\bigr)^\top \in \mathbb{R}^{n\times r}$.
We then consider the fusion-regularized objective:
\begin{equation}\label{obj0}
	\begin{aligned} 
		Q (\bm{\beta},\bm{\zeta},\bm{B},\bm{Z};\bm{A})&=L(\bm{\beta},\bm{\zeta};\bm{A})+J_1(\bm{\beta}, \bm{B})+ J_2(\bm{\zeta}, \bm{Z}) +J_3(\bm{B})
			+J_4(\bm{Z}),
	\end{aligned} 
\end{equation}
where
\begin{equation}\nonumber
	\begin{aligned} 
		& J_1(\bm{\beta}, \bm{B}) = \lambda_1 \|\bm{\beta}_{-\{0\}}-\bm{B}\|_F^2, ~~~~~~~~
        J_2(\bm{\zeta}, \bm{Z}) = \lambda_2\|\bm{\zeta}_{-\{n+1\}}-\bm{Z}\|_F^2, \\
        & J_3(\bm{B}) = \sum_{1\leqslant i<j\leqslant n}\rho\bigl(\|\bm{b}_i-\bm{b}_j\|_2;\lambda_3\bigr), \
        J_4(\bm{Z}) = \sum_{1\leqslant i<j\leqslant n}\rho\bigl(\|\bm{z}_i-\bm{z}_j\|_2;\lambda_4\bigr),
	\end{aligned} 
\end{equation}
$\bm{\beta}_{-\{0\}}$ is $\bm{\beta}$ with $\bm{\beta}_0$ removed, $\bm{\zeta}_{-\{n+1\}}$ is $\bm{\zeta}$ with $\bm{\zeta}_{n+1}$ removed, and $\rho(\cdot;\lambda)$ denotes the MCP fusion penalty with regularization parameter $\lambda$. The first two quadratic terms shrink the embeddings toward their node-specific centers, while the last two fusion terms encourage similar centers to merge. Consequently, the numbers of distinct rows in $\bm{B}$ and $\bm{Z}$ among real nodes yield the estimated numbers of tail and head communities. More precisely, let $\{\widehat{\bm{\mu}}_1,\ldots,\widehat{\bm{\mu}}_{\widehat K_{\mathrm{ta}}}\}$ be the distinct values among $\{\widehat{\bm{b}}_1,\ldots,\widehat{\bm{b}}_n\}$, and let $\{\widehat{\bm{\nu}}_1,\ldots,\widehat{\bm{\nu}}_{\widehat K_{\mathrm{he}}}\}$ be the distinct values among $\{\widehat{\bm{z}}_1,\ldots,\widehat{\bm{z}}_n\}$. Then the estimated community assignment functions are defined by $\widehat{\psi}_{\mathrm{ta}}(i)=k$ if $\widehat{\bm{b}}_i=\widehat{\bm{\mu}}_k$ and $\widehat{\psi}_{\mathrm{he}}(j)=l$ if $\widehat{\bm{z}}_j=\widehat{\bm{\nu}}_l.$

The objective function in \eqref{obj0} balances fit to the observed hyperedges against a parsimonious community structure. The penalty terms describe each node through a tail-side center and a head-side center whose distinct values are determined by the data.

The quadratic penalties $J_1$ and $J_2$ separate node-specific interaction profiles, which the likelihood estimates, from community membership, which the fusion penalties determine. This differs from post-hoc clustering, which carries embedding estimation error directly into the clustering decisions. The MCP fusion penalties $J_3$ and $J_4$ fuse selectively: small differences between centers are penalized strongly so that similar centers merge, while large differences are left essentially untouched. The fusion penalties thus determine both the number and the composition of communities. The null nodes $0$ and $n+1$ are excluded from all penalty terms in \eqref{obj0}. Their embeddings $\bm{\beta}_0$ and $\bm{\zeta}_{n+1}$ are fixed at the normalized unit vector and do not appear in $J_1,\dots, J_4$. Consequently, each null node can be viewed as a singleton community in its respective side: the tail-null node $0$ forms its own tail community, and the head-null node $n+1$ forms its own head community. This treatment is natural, as the null nodes are artificial constructs introduced for tensor completion and should not influence community detection among real nodes.

The combination of quadratic shrinkage toward node-specific centers and pairwise fusion of the centers was introduced by \citet{ren2023consistent} for a single embedding in undirected networks, and \citet{zhang2025network} applied it to the out- and in-node embeddings of directed networks. In \eqref{obj0}, the tail-side terms $J_1$ and $J_3$ and the head-side terms $J_2$ and $J_4$ are tied together by one likelihood over hyperedges of different sizes, in which every hyperedge involves several tail embeddings and several head embeddings at once. The two partitions are therefore estimated jointly, and their numbers of communities are allowed to differ. Table~\ref{tab:related} compares FOLD with these methods in setting and theoretical guarantees.

\begin{table}[ht]
    \centering
    \caption{Setting and theoretical guarantees of FOLD and of the most closely related embedding-based methods. Check marks indicate covered features. Global recovery means exact recovery of the partitions with unknown numbers of communities under a guarantee that does not depend on the choice of a local minimizer.}
    \label{tab:related}
    \small
    \setlength{\tabcolsep}{4.5pt}
    \begin{tabular}{@{}lcccccc@{}}
        \toprule
         & \multicolumn{3}{c}{Setting} & \multicolumn{3}{c}{Theory}\\
        \cmidrule(lr){2-4}\cmidrule(l){5-7}
        Method & \shortstack{Higher-order\\interactions} & \shortstack{Directed\\roles} & \shortstack{Unknown\\numbers} & \shortstack{Global\\recovery} & \shortstack{Minimax\\lower bound} & \shortstack{Necessary\\condition}\\
        \midrule
        \citet{zhen2023community} & \checkmark & -- & -- & -- & -- & --\\
        \citet{zhang2022directed} & -- & \checkmark & -- & -- & -- & --\\
        \citet{ren2023consistent} & -- & -- & \checkmark & -- & -- & --\\
        \citet{zhang2025network} & -- & \checkmark & \checkmark & -- & -- & --\\
        FOLD & \checkmark & \checkmark & \checkmark & \checkmark & \checkmark & \checkmark\\
        \bottomrule
    \end{tabular}
\end{table}

The tail and head sides are treated asymmetrically: there is no constraint that the two community structures coincide. This respects the directional nature of the data and lets the source and target roles of a node differ, while the shared likelihood couples them. 

The objective in \eqref{obj0} is nonconvex because of the likelihood term and the MCP penalties. Its penalties, however, separate into quadratic terms that couple embeddings to centers and fusion terms that couple centers to each other. This structure leads to an alternating algorithm with tractable subproblems (Section B.1 of the Supplementary Material).

\section{Statistical properties}\label{sec:theory}
In this section, we establish some theoretical properties of the proposed method. For fixed integers $p$ and $q$ and a positive constant $c_0 \geq 1$, let $\Omega_{\Theta}$ denote the parameter space:
\begin{equation*}
    \begin{aligned}
        \Omega_{\Theta}= \Bigl\{ \bm{\Theta}: &\; \bm{\Theta}=\mathcal{I}\times_1 \bm{\beta}\times_2 \cdots\times_p \bm{\beta}\times_{p+1} \bm{\zeta}\times_{p+2} \cdots\times_{p+q} \bm{\zeta}, \bm{\beta}_0=r^{-1/2}\bm{1}_r,\\
        &  \bm{\zeta}_{n+1}=r^{-1/2} \bm{1}_r, \max _{i\in[n]\cup\{0\}}\max_{j\in[n+1]} \{ \|\bm{\beta}_i\|_{2}, \|\bm{\zeta}_j\|_{2} \} \leqslant c_{0} \Bigr\} \subset \mathbb{R}^{(n+1) \times \cdots \times(n+1)}.
    \end{aligned}
\end{equation*}
It is clear that $\Omega_{\Theta}$ is a compact subset of $\mathbb{R}^{(n+1) \times \cdots \times(n+1)}$, and for any $\bm{\Theta}\in \Omega_{\Theta}$, its rank is at most $r$, which is allowed to diverge. Throughout, \eqref{obj0} and the other objective functions below are minimized in the parameter space $\Omega_{\Theta}$, and a local minimizer of \eqref{obj0} refers to a local minimizer of this constrained problem.

Denote the true parameters as $\bm{\beta}^{*}=\bigl(\bm{\beta}_0^{*},\bm{\beta}_1^{*},\ldots,\bm{\beta}_{n}^{*}\bigr)^\top \in \mathbb{R}^{(n+1)\times r}$, $\bm{\zeta}^{*}=\bigl(\bm{\zeta}_1^{*},\bm{\zeta}_2^{*},\ldots,\bm{\zeta}_{n+1}^{*}\bigr)^\top \in \mathbb{R}^{(n+1)\times r}$, $\boldsymbol{B}^{*} = ( \boldsymbol{b}^{*}_1, \cdots, \boldsymbol{b}^{*}_n )^{\top}$, $\boldsymbol{Z}^{*} = ( \boldsymbol{z}^{*}_1, \cdots, \boldsymbol{z}^{*}_n )^{\top}$, and $\bm{\Theta}^{*}=\mathcal{I}\times_1 \bm{\beta}^{*} \times_2 \cdots\times_p \bm{\beta}^{*} \times_{p+1} \bm{\zeta}^{*} \times_{p+2} \cdots\times_{p+q} \bm{\zeta}^{*}.$ Assume that $\bm{\Theta}^{*} \in \Omega_{\Theta}$.
Let $\{ \boldsymbol{\mu}^{*}_1, \cdots, \boldsymbol{\mu}^{*}_{K^*_{\mathrm{ta}}} \} $ and $\{ \boldsymbol{\nu}^{*}_1, \cdots, \boldsymbol{\nu}^{*}_{K^*_{\mathrm{he}}} \} $ be the distinct values of  $\{\boldsymbol{b}^{*}_1, \cdots, \boldsymbol{b}^{*}_n \}$ and $\{ \boldsymbol{z}^{*}_1, \cdots, \boldsymbol{z}^{*}_n \}$, respectively, with the true numbers of tail and head communities $K^*_{\mathrm{ta}}, K^*_{\mathrm{he}} \geqslant 2$. Let $\psi_{\mathrm{ta}}^*:[n]\to[K_{\mathrm{ta}}^*],\psi_{\mathrm{he}}^*:[n]\to[K_{\mathrm{he}}^*]$ denote the true tail- and head-community assignment functions. Then,  $\mathcal{T}^{*}_{k} = \{i: \psi^*_{\mathrm{ta}}(i) = k, i=1, \cdots, n \}$ for $k=1, \cdots, K^*_{\mathrm{ta}}$ and $\mathcal{H}^{*}_{k} = \{i: \psi^*_{\mathrm{he}}(i) = k , i=1, \cdots, n \}$ for $k=1, \cdots, K^*_{\mathrm{he}}$ are the true tail and head community structures, respectively. Then $\boldsymbol{\mu}_k^*=\frac{1}{\left|\mathcal{T}_k^*\right|} \sum_{i \in \mathcal{T}_k^*} \boldsymbol{\beta}_i^*$, $\boldsymbol{\nu}_k^*=\frac{1}{\left|\mathcal{H}_k^*\right|} \sum_{i \in \mathcal{H}_k^*} \boldsymbol{\zeta}_i^*$,
$\boldsymbol{b}_i^*=\boldsymbol{\mu}_k^*, \text { for } i \in \mathcal{T}_k^*$,  $\boldsymbol{z}_i^*=\boldsymbol{\nu}_k^*, \text { for } i \in \mathcal{H}_k^* $.
Theorems \ref{Theo1} and \ref{Theo2} and Corollary \ref{Cor1}, together with the lower bound in Theorem \ref{Theo4} and the necessary condition in Theorem \ref{Theo5}, hold for every $p+q\geqslant 2$ and thus cover ordinary directed networks, which correspond to $p=q=1$.
	
Let $\varrho_{\omega}^*=\frac{s_n}{1+\exp(-\theta_{\omega}^*)}$ and $\varrho_{\omega}=\frac{s_n}{1+\exp(-\theta_{\omega})}$ with $\omega=(i_1,\ldots,i_p; j_1,\ldots,j_q)\in \Omega_{p,q}^{\mathrm{ord}}$. Define the averaged squared Hellinger distance between $\bm{\Theta}$ and $\bm{\Theta}^{*}$ as:
\begin{equation}\notag
    h^{2}\left(\bm{\Theta}, \bm{\Theta}^{*}\right)=\frac{1}{\phi(n,p,q)} \sum_{\omega\in \Omega_{p,q}^{\mathrm{ord}}} \left[\left(\varrho_\omega^{1/2}-\left(\varrho_\omega^{*}\right)^{1 / 2}\right)^{2}+\left((1-\varrho_\omega)^{1 / 2}-\left(1-\varrho^{*}_\omega\right)^{1 / 2}\right)^{2}\right].
\end{equation}
Let $\|\bm{\Theta}-\bm{\Theta}^{*}\|_{\Omega}=\bigl\{\sum_{\omega\in\Omega_{p,q}^{\mathrm{ord}}}(\theta_\omega-\theta_\omega^{*})^2\bigr\}^{1/2}$ denote the Frobenius norm of $\bm{\Theta}-\bm{\Theta}^{*}$ restricted to the legal canonical tuples. When $\{ \mathcal{T}^{*}_{k} \}_{k=1}^{K^*_{\mathrm{ta}}}$ and $\{ \mathcal{H}^{*}_{k} \}_{k=1}^{K^*_{\mathrm{he}}}$ are known, define $\mathcal{B}_{\mathcal{T}^{*}} = \{ \boldsymbol{B} \in \mathbb{R}^{n \times r}: \boldsymbol{b}_i = \boldsymbol{b}_j, \text{ for any } i,j \in \mathcal{T}^{*}_{k} \ \mbox{and} \ k = 1, \cdots, K^*_{\mathrm{ta}} \}$ and $\mathcal{Z}_{\mathcal{H}^{*}} = \{ \boldsymbol{Z} \in \mathbb{R}^{n \times r}: \boldsymbol{z}_i = \boldsymbol{z}_j, \text{ for any } i,j \in \mathcal{H}^{*}_{k} \ \mbox{and} \ k = 1, \cdots, K^*_{\mathrm{he}} \}$. Let $(\widehat{\bm{\beta}}^{o}, \widehat{\bm{\zeta}}^{o}, \widehat{\bm{B}}^{o}, \widehat{\bm{Z}}^{o})$ be a solution of $\argmin_{ \bm{\beta}, \bm{\zeta}, \bm{B}\in \mathcal{B}_{\mathcal{T}^{*}}, \bm{Z}\in \mathcal{Z}_{\mathcal{H}^{*}}} \{ L(\bm{\beta},\bm{\zeta};\bm{A})+J_1(\bm{\beta}, \bm{B})+ J_2(\bm{\zeta}, \bm{Z})\}$ in the parameter space $\Omega_{\Theta}$. Define $\widehat{\bm{\Theta}}^{o}=\mathcal{I}\times_1 \widehat{\bm{\beta}}^{o} \times_2 \cdots\times_p \widehat{\bm{\beta}}^{o} \times_{p+1} \widehat{\bm{\zeta}}^{o} \times_{p+2} \cdots\times_{p+q} \widehat{\bm{\zeta}}^{o}$.
\begin{Mth}\label{Theo1}
	Assume that $J_1(\bm{\beta}^{*}, \bm{B}^{*}) + J_2(\bm{\zeta}^{*}, \bm{Z}^{*}) \leqslant \frac{c_1  r \log n}{2n^{p+q-1}}$ for some $c_1 > 0$. Then there exist some constants $c_{2}, c_{3}>0$, such that when $n$ is sufficiently large,
	\begin{equation}\nonumber
		\begin{aligned}
			\pr\Bigl(h(\widehat{\bm{\Theta}}^o,\bm{\Theta}^*) \leqslant \sqrt{\frac{c_3  r \log n}{n^{p+q-1}} } \Bigr) & \geqslant 1 - 2\exp\bigl(-c_2n r \log n\bigr), \\
				\pr\Bigl( \frac{1}{\phi^{1/2}(n,p,q)}\|\widehat{\bm{\Theta}}^o-\bm{\Theta}^*\|_{\Omega} \leqslant \sqrt{\frac{c_3  r \log n}{s_n n^{p+q-1}} }\Bigr) & \geqslant 1 - 2\exp\bigl(-c_2 n r \log n\bigr).
		\end{aligned}
	\end{equation}
\end{Mth}
Theorem \ref{Theo1} bounds the errors of the oracle estimator $\widehat{\bm{\Theta}}^o$, which knows the true tail and head community structures in advance, under the averaged Hellinger distance and the Frobenius norm over the legal canonical tuples. Both errors vanish whenever $r\log n/(s_n n^{p+q-1})\to 0$. Coordinates of $\bm{\Theta}$ with $\delta_\omega=0$ do not correspond to any directed hyperedge and do not enter the likelihood, and every legal coordinate equals the coordinate of its canonical representative. Both error measures are therefore taken over $\Omega_{p,q}^{\mathrm{ord}}$. The convergence rates are governed by the latent dimension $r$, the network size $n$, the order parameters $p,q$, and the sparsity factor $s_n$. In particular, the Frobenius norm rate involves an extra factor $s_n^{-1/2}$, reflecting the difficulty of estimating the transformed interaction parameter $\theta_{\omega}$ when the hypergraph is sparse. The rates are of the same order as those of \citet{zhen2023community} for undirected hypergraphs, with the tensor order replaced by $p+q$. Our proof works directly with the legal canonical tuples, on which the directed likelihood is defined, and bounds likelihood ratios through Hellinger distances and a finite net of $\Omega_{\Theta}$ (Section A.2 of the Supplementary Material). It uses the true partitions only through the fact that $(\bm{B}^*,\bm{Z}^*)$ satisfies the constraint of the minimization, so the two bounds hold, with the same constants, for every minimizer of $L(\bm{\beta},\bm{\zeta};\bm{A})+J_1(\bm{\beta},\bm{B})+J_2(\bm{\zeta},\bm{Z})$ over $\Omega_{\Theta}$ under any constraint on $(\bm{B},\bm{Z})$ that $(\bm{B}^*,\bm{Z}^*)$ satisfies. In particular, the maximum likelihood estimator over $\Omega_{\Theta}$, which corresponds to $\lambda_1=\lambda_2=0$ and no constraint, attains the same rates without knowledge of the partitions. In order to establish the theoretical properties of community detection, additional conditions and notation are needed.
\begin{Con}\label{balan}
	There exists a positive constant $C_0$ such that $\frac{K_{\mathrm{ta}}^*\min_{1\leqslant k\leqslant K_{\mathrm{ta}}^*}|\mathcal{T}_k^*|}{n}\geqslant C_0$ and $\frac{K_{\mathrm{he}}^*\min_{1\leqslant l\leqslant K_{\mathrm{he}}^*}|\mathcal{H}_l^*|}{n}\geqslant C_0$.
\end{Con}
Condition \ref{balan} indicates that all the true tail and head communities will not degenerate asymptotically and allows the number of communities to diverge. A similar condition has been imposed in the existing network community detection literature \citep{Ke2019Community,zhang2022directed}. 
Let $\bm{M}^*=\bigl(\bm{\mu}_1^*,\ldots,\bm{\mu}_{K_{\mathrm{ta}}^*}^*\bigr)^\top$ and $\bm{N}^*=\bigl(\bm{\nu}_1^*,\ldots,\bm{\nu}_{K_{\mathrm{he}}^*}^*\bigr)^\top$, and define
$\bm{W}^*
	=
	\mathcal{I}
	\times_1\bm{M}^*
	\times_2\cdots\times_p\bm{M}^*
	\times_{p+1}\bm{N}^*
	\times_{p+2}\cdots\times_{p+q}\bm{N}^*.$
For $k\in[K_{\mathrm{ta}}^*]$, let $\bm{W}_{\mathrm{ta},k}^*$ be the $(p+q-1)$-order subtensor obtained by fixing the first tail index of $\bm{W}^*$ at $k$, namely, $[\bm{W}_{\mathrm{ta},k}^*]_{k_2,\ldots,k_p,l_1,\ldots,l_q}
	=
	[\bm{W}^*]_{k,k_2,\ldots,k_p,l_1,\ldots,l_q}$. Similarly, for $l\in[K_{\mathrm{he}}^*]$, let $\bm{W}_{\mathrm{he},l}^*$ be the $(p+q-1)$-order subtensor obtained by fixing the first head index of $\bm{W}^*$ at $l$, namely, $[\bm{W}_{\mathrm{he},l}^*]_{k_1,\ldots,k_p,l_2,\ldots,l_q}
	=
	[\bm{W}^*]_{k_1,\ldots,k_p,l,l_2,\ldots,l_q}$. The tensor $\bm{W}^*$ is the community-level representation of the node-level interaction tensor $\bm{\Theta}^*$ rather than an additional population parameter. By multilinearity and the definitions of $\bm{\mu}_k^*$ and $\bm{\nu}_l^*$, for $k_1,\ldots,k_p\in[K_{\mathrm{ta}}^*]$ and $l_1,\ldots,l_q\in[K_{\mathrm{he}}^*]$, $[\bm{W}^*]_{k_1,\ldots,k_p,l_1,\ldots,l_q}
=
\frac{1}{\prod_{u=1}^p|\mathcal{T}_{k_u}^*|\prod_{v=1}^q|\mathcal{H}_{l_v}^*|}
\sum_{\substack{i_1\in\mathcal{T}_{k_1}^*,\ldots,i_p\in\mathcal{T}_{k_p}^*\\
j_1\in\mathcal{H}_{l_1}^*,\ldots,j_q\in\mathcal{H}_{l_q}^*}}
[\bm{\Theta}^*]_{i_1,\ldots,i_p,j_1,\ldots,j_q}.$
Hence, $\bm{W}^*$ is the blockwise mean of $\bm{\Theta}^*$ under the true tail and head partitions: it retains the directed interaction pattern explained by community membership while averaging the node-specific variation within each community. Accordingly, $\bm{W}_{\mathrm{ta},k}^*$ and $\bm{W}_{\mathrm{he},l}^*$ describe the complete tail-side and head-side interaction profiles of communities $k$ and $l$, respectively. Define $\Delta_n
	=
	\min\left\{
	\frac{\min_{k\neq k'}\|\bm{W}_{\mathrm{ta},k}^*-\bm{W}_{\mathrm{ta},k'}^*\|_F}
	{\{(K_{\mathrm{ta}}^*)^{p-1}(K_{\mathrm{he}}^*)^q\}^{1/2}},
	\,
	\frac{\min_{l\neq l'}\|\bm{W}_{\mathrm{he},l}^*-\bm{W}_{\mathrm{he},l'}^*\|_F}
	{\{(K_{\mathrm{ta}}^*)^p(K_{\mathrm{he}}^*)^{q-1}\}^{1/2}}
	\right\}$. The quantity $\Delta_n$ is the smallest normalized Frobenius separation between two such profiles. 

\begin{Con}\label{profile_sep}
	There exists a sufficiently large positive constant $C>1$ such that
	$\Delta_n^2
		>
		C\max\{\lambda_3^2,\lambda_4^2\}
		>
		\min\{\lambda_3^2,\lambda_4^2\}
		>
		C\Bigl\{\frac{(K_{\mathrm{ta}}^*\vee K_{\mathrm{he}}^*)r\log n}{s_n n^{p+q-1}}+\frac{(K_{\mathrm{ta}}^*\vee K_{\mathrm{he}}^*)^3}{n^2}\Bigr\}$.
\end{Con}

Condition \ref{profile_sep} requires this between-community signal to dominate both the fusion scales and the estimation error of the community-level tensor $\bm{W}^*$. The estimation error has two parts. The term $\frac{(K_{\mathrm{ta}}^*\vee K_{\mathrm{he}}^*)r\log n}{s_n n^{p+q-1}}$ is the stochastic estimation error over the legal canonical tuples given in Theorem \ref{Theo1}. The term $\frac{(K_{\mathrm{ta}}^*\vee K_{\mathrm{he}}^*)^3}{n^2}$ accounts for the index tuples in which a node appears more than once. These tuples are not legal, but they enter the blockwise mean that defines $\bm{W}^*$, and their proportion in each community block is of order $(K_{\mathrm{ta}}^*\vee K_{\mathrm{he}}^*)/n$. Together with the boundedness condition, profile separation implies separation of the center vectors. Such a separation condition is common in the network community-detection literature \citep{zhang2022directed,zhen2023community}.
The lower bound on $\min\{\lambda_3^2,\lambda_4^2\}$ ensures that the fusion penalty is strong enough to shrink within-community centers together. The choice of $\lambda_1$ and $\lambda_2$ specified in Theorem~\ref{Theo2} ensures that the quadratic shrinkage does not dominate the likelihood. Because $\Delta_n\leqslant 2c_0^{p+q}$, Condition \ref{profile_sep} can hold only when $\frac{(K_{\mathrm{ta}}^*\vee K_{\mathrm{he}}^*)r\log n}{s_n n^{p+q-1}}$ is bounded by a small constant. When $r$ and $K_{\mathrm{ta}}^*\vee K_{\mathrm{he}}^*$ are of the constant order, the sparsity factor must therefore satisfy $s_n\geqslant c\log n/n^{p+q-1}$ for a sufficiently large constant $c$, which matches the minimal requirement for consistent estimation in hypergraph models \citep{zhen2023community}.

\begin{Mth}\label{Theo2}
	Suppose Conditions \ref{balan} and \ref{profile_sep} hold. Assume that $0<\lambda_1,\lambda_2\leqslant\frac{C_1 r\log n}{n^{p+q}}$ for a positive constant $C_1$. Then, when $n$ is sufficiently large, $(\widehat{\bm{\beta}}^{o}, \widehat{\bm{\zeta}}^{o}, \widehat{\bm{B}}^{o}, \widehat{\bm{Z}}^{o})$ is a local minimizer of (\ref{obj0}) with probability at least $1 - 2\exp\bigl(-c_2 n r \log n\bigr) $, where $c_2$ is the constant in Theorem \ref{Theo1} corresponding to $c_1=16c_0^2C_1$.
\end{Mth}
Theorem \ref{Theo2} establishes that, under appropriate separation and sparsity conditions, the oracle estimator is actually a local minimizer of the original fusion-regularized objective function with high probability. Because the embeddings are bounded, this choice of $\lambda_1$ and $\lambda_2$ implies the assumption of Theorem \ref{Theo1} with $c_1=16c_0^2C_1$. The conditions of Theorem \ref{Theo2} involve the true embeddings only through the community-level tensor $\bm{W}^*$ and the boundedness of $\Omega_{\Theta}$, and the node embeddings within a community need not coincide. Theorem \ref{Theo2} is local: it concerns the objective function \eqref{obj0} only in a neighborhood of the oracle estimator, whose fused centers reproduce the true tail and head partitions. A more comprehensive insight into the penalized fusion-based regularized negative log-likelihood function in \eqref{obj0} can be gained by combining Theorems \ref{Theo1} and \ref{Theo2}, as summarized in the following corollary.
\begin{Cor}\label{Cor1}
	Suppose all the conditions in Theorem \ref{Theo2} hold. Then, with probability approaching 1, there exists a local minimizer of (\ref{obj0}), denoted by $(\widehat{\bm{\beta}}, \widehat{\bm{\zeta}}, \widehat{\bm{B}}, \widehat{\bm{Z}})$, and $\widehat{\bm{\Theta}}=\mathcal{I}\times_1 \widehat{\bm{\beta}} \times_2 \cdots\times_p \widehat{\bm{\beta}} \times_{p+1} \widehat{\bm{\zeta}} \times_{p+2} \cdots\times_{p+q} \widehat{\bm{\zeta}}$, $\widehat{\mathcal T}_k=\{i:\widehat\psi_{\mathrm{ta}}(i)=k\},\widehat{\mathcal H}_k=\{i:\widehat\psi_{\mathrm{he}}(i)=k\}$, after aligning the community labels through an appropriate permutation, such that: \\
	(1) $\pr\bigl( \widehat{K}_{\mathrm{ta}} = K_{\mathrm{ta}}^* \bigr) \rightarrow 1, \quad \pr\bigl( \widehat{K}_{\mathrm{he}} = K_{\mathrm{he}}^* \bigr) \rightarrow 1$; \\
	(2) $\pr\bigl( \widehat{\mathcal{T}}_k = \mathcal{T}_k^*,\ \text{for } k = 1,\dots,\widehat{K}_{\mathrm{ta}} \bigr) \rightarrow 1, \quad \pr\bigl( \widehat{\mathcal{H}}_l = \mathcal{H}_l^*,\ \text{for } l = 1,\dots,\widehat{K}_{\mathrm{he}} \bigr) \rightarrow 1$; \\
	(3) $h ( \widehat{\boldsymbol{\Theta}}, \boldsymbol{\Theta}^* ) = O_p ( \sqrt{\frac{  r \log n}{n^{p+q-1}} } )$;
	(4) $\frac{1}{\phi^{1/2}(n,p,q)}\|\widehat{\bm{\Theta}}-\bm{\Theta}^*\|_{\Omega} = O_p ( \sqrt{\frac{ r \log n}{s_n n^{p+q-1}} } )$. 
\end{Cor}
Corollary 1 combines Theorems 1 and 2 to deliver the main practical message of the proposed method, that is, there exists a local minimizer of the fusion-regularized objective function \eqref{obj0} that simultaneously achieves consistent estimation of the number of tail communities, the number of head communities, the exact community memberships, and the underlying probability tensor. Notably, the consistency of community structures, especially the numbers of communities $\widehat K_{\mathrm{ta}}$ and $\widehat K_{\mathrm{he}}$, is established, in contrast to most existing spectral or likelihood-based methods for network or hypergraph community detection requiring the true numbers to be known a priori. 
Such consistency of the numbers of communities can be difficult to establish in the hypergraph embedding model. The popular CV methods \citep{chen2018network,li2020edge} can only guarantee that the number of communities can be consistently overestimated. The consistent estimation result in the StGoF method \citep{jin2023optimal} heavily relies on the restrictive DCBM framework for ordinary networks. Moreover, by Theorem \ref{Theo4} below, the convergence rates in parts (3) and (4) are minimax optimal up to the factor $\log n$. Note that when $p+q=2$, the rate $\sqrt{\frac{r\log n}{n}}$ aligns with the theoretical results for directed networks that only consider a single head and tail node \citep{zhang2022directed}. Corollary \ref{Cor1} extends the consistency results of \citet{zhen2023community} for undirected hypergraphs, which assume a known number of communities, and it parallels the local-minimizer results of \citet{ren2023consistent} and \citet{zhang2025network} for ordinary networks. Two features of the present setting require additional arguments. First, the separation $\Delta_n$ is defined through two families of interaction profiles, one for each side, with different normalizations. Second, the community-level tensor $\bm{W}^*$ averages over index tuples that are not legal, and their contribution is bounded explicitly, which yields the second term in Condition \ref{profile_sep}.
\begin{Rem}\label{rem:local}
	Theorem \ref{Theo2} and Corollary \ref{Cor1} concern one particular local minimizer of \eqref{obj0}. The objective function in \eqref{obj0} is not convex, and under the conditions of Theorem \ref{Theo2} it has other local minimizers. For example, let $(\widehat{\bm{\beta}}^{f},\widehat{\bm{\zeta}}^{f},\widehat{\bm{B}}^{f},\widehat{\bm{Z}}^{f})$ be a minimizer of $L(\bm{\beta},\bm{\zeta};\bm{A})+J_1(\bm{\beta},\bm{B})+J_2(\bm{\zeta},\bm{Z})$ in the parameter space $\Omega_{\Theta}$ under the constraint that all rows of $\bm{B}$ coincide and all rows of $\bm{Z}$ coincide. When $n$ is sufficiently large, this completely fused estimator is also a local minimizer of \eqref{obj0}, and on the event of Theorem \ref{Theo2} its objective value is smaller than that of the oracle estimator. Both facts are proved in Section A.6 of the Supplementary Material.
\end{Rem}
We next give a guarantee that does not depend on the choice of a local minimizer. Let $(\widehat{\bm{\beta}}^{\mathrm{ml}},\widehat{\bm{\zeta}}^{\mathrm{ml}})$ be a minimizer of $L(\bm{\beta},\bm{\zeta};\bm{A})$ in the parameter space $\Omega_{\Theta}$, and let $\widehat{\bm{\Theta}}^{\mathrm{ml}}=\mathcal{I}\times_1 \widehat{\bm{\beta}}^{\mathrm{ml}} \times_2 \cdots\times_p \widehat{\bm{\beta}}^{\mathrm{ml}} \times_{p+1} \widehat{\bm{\zeta}}^{\mathrm{ml}} \times_{p+2} \cdots\times_{p+q} \widehat{\bm{\zeta}}^{\mathrm{ml}}$ be the maximum likelihood estimator of $\bm{\Theta}^*$, with entries $\widehat{\theta}^{\mathrm{ml}}_\omega$. For two distinct nodes $i,i'\in[n]$, let $\mathcal{C}^{\mathrm{ta}}_{ii'}$ be the set of tuples in $\Omega_{p,q}^{\mathrm{ord}}$ whose tail contains $i$ and which do not contain $i'$. For $\omega\in\mathcal{C}^{\mathrm{ta}}_{ii'}$, let $\omega^{i\to i'}$ be the tuple in $\Omega_{p,q}^{\mathrm{ord}}$ obtained from $\omega$ by replacing the tail node $i$ with $i'$. We define the tail slice distance between $i$ and $i'$ as
\begin{equation}\label{slice-ta}
    \widehat{d}_{\mathrm{ta}}(i,i')=\Bigl\{\frac{1}{|\mathcal{C}^{\mathrm{ta}}_{ii'}|}\sum_{\omega\in\mathcal{C}^{\mathrm{ta}}_{ii'}}\bigl(\widehat{\theta}^{\mathrm{ml}}_{\omega}-\widehat{\theta}^{\mathrm{ml}}_{\omega^{i\to i'}}\bigr)^2\Bigr\}^{1/2}.
\end{equation}
The tail slice distance measures how much the fitted interaction parameters change when $i'$ takes the place of $i$ as a tail node, averaged over all tuples in which this replacement is possible. It is small when $i$ and $i'$ have similar tail-side interaction profiles. The head slice distance $\widehat{d}_{\mathrm{he}}(j,j')$ is defined in the same way, with the set $\mathcal{C}^{\mathrm{he}}_{jj'}$ of tuples in $\Omega_{p,q}^{\mathrm{ord}}$ whose head contains $j$ and which do not contain $j'$, and with $j'$ replacing $j$ in the head. Given a threshold $\tau_n>0$, we join two nodes $i$ and $i'$ whenever $\widehat{d}_{\mathrm{ta}}(i,i')\leqslant\tau_n$, and we let $\widetilde{\mathcal{T}}_1,\ldots,\widetilde{\mathcal{T}}_{\widetilde{K}_{\mathrm{ta}}}$ be the connected components of the resulting graph on $[n]$. The sets $\widetilde{\mathcal{H}}_1,\ldots,\widetilde{\mathcal{H}}_{\widetilde{K}_{\mathrm{he}}}$ are obtained in the same way from $\widehat{d}_{\mathrm{he}}$. The procedure does not take the numbers of communities as input.
\begin{Con}\label{cohesion}
	Let $\varsigma_n=\max_{i\in[n]}\{\|\bm{\beta}_i^*-\bm{b}_i^*\|_2\vee\|\bm{\zeta}_i^*-\bm{z}_i^*\|_2\}$ denote the within-community variation. There exists a sufficiently small positive constant $c_\varsigma$ such that $\varsigma_n\leqslant c_\varsigma\Delta_n$.
\end{Con}
Condition \ref{cohesion} requires the within-community variation to be small relative to the separation between communities, and it does not require this variation to vanish. Under this condition, two nodes in the same community have closer interaction profiles than two nodes in different communities, so the true partitions are obtained by grouping nodes whose profiles differ by less than a threshold lying between the within-community variation and the separation. Theorem \ref{Theo3} makes this precise.
\begin{Mth}\label{Theo3}
	Suppose Conditions \ref{balan} and \ref{cohesion} hold, and that $\frac{K_{\mathrm{ta}}^*\vee K_{\mathrm{he}}^*}{n}$ and $\frac{r\log n}{s_n n^{p+q-2}\Delta_n^2}$ are bounded by a sufficiently small positive constant. Let $c_d=\{C_0^{p+q-1}/(2pq\,p!\,q!)\}^{1/2}$ and let the threshold satisfy $c_d\Delta_n/4\leqslant\tau_n<3c_d\Delta_n/4$. Then, when $n$ is sufficiently large, with probability at least $1-\exp\{-2(p+q)nr\log n\}$, we have $\widetilde{K}_{\mathrm{ta}}=K_{\mathrm{ta}}^*$, $\widetilde{K}_{\mathrm{he}}=K_{\mathrm{he}}^*$, and, after relabeling, $\widetilde{\mathcal{T}}_k=\mathcal{T}_k^*$ for $k=1,\ldots,K_{\mathrm{ta}}^*$ and $\widetilde{\mathcal{H}}_l=\mathcal{H}_l^*$ for $l=1,\ldots,K_{\mathrm{he}}^*$.
\end{Mth}
Theorem \ref{Theo3} holds for every maximum likelihood estimator and does not involve the penalized objective \eqref{obj0}. The admissible range of $\tau_n$ plays the same role as the admissible range of $\lambda_3$ and $\lambda_4$ in Condition \ref{profile_sep}. When $r$ is of the constant order and $\Delta_n$ is bounded away from zero, the condition of Theorem \ref{Theo3} reads $s_n\geqslant c\log n/n^{p+q-2}$ for a sufficiently large constant $c$, which allows sparse hypergraphs whenever $p+q\geqslant 3$.

The partitions of Theorem \ref{Theo3} also locate the local minimizer of Theorem \ref{Theo2} from the data. Let $(\widehat{\bm{\beta}}^{\mathrm{rf}},\widehat{\bm{\zeta}}^{\mathrm{rf}},\widehat{\bm{B}}^{\mathrm{rf}},\widehat{\bm{Z}}^{\mathrm{rf}})$ be a minimizer of $L(\bm{\beta},\bm{\zeta};\bm{A})+J_1(\bm{\beta},\bm{B})+J_2(\bm{\zeta},\bm{Z})$ in the parameter space $\Omega_{\Theta}$ under the constraint that the rows of $\bm{B}$ coincide within each $\widetilde{\mathcal{T}}_k$ and the rows of $\bm{Z}$ coincide within each $\widetilde{\mathcal{H}}_l$, and let $\widehat{\bm{\Theta}}^{\mathrm{rf}}$ be the corresponding tensor. We call it the refitted estimator.
\begin{Cor}\label{Cor2}
	Suppose that the conditions in Theorem \ref{Theo3} hold and that $0<\lambda_1,\lambda_2\leqslant\frac{C_1 r\log n}{n^{p+q}}$ for a positive constant $C_1$. Then, when $n$ is sufficiently large, with probability at least $1-2\exp\{-2(p+q)nr\log n\}$, the refitted estimator is a solution of the minimization problem that defines the oracle estimator, and $\widehat{\bm{\Theta}}^{\mathrm{rf}}$ satisfies the two bounds of Theorem \ref{Theo1}. If, in addition, Condition \ref{profile_sep} holds, then on this event the refitted estimator is a local minimizer of \eqref{obj0}, and the distinct rows of $\widehat{\bm{B}}^{\mathrm{rf}}$ and $\widehat{\bm{Z}}^{\mathrm{rf}}$ determine exactly the true tail and head partitions.
\end{Cor}
Corollary \ref{Cor1} asserts the existence of a local minimizer with the desired properties. Corollary \ref{Cor2} shows that, given a threshold in the range of Theorem \ref{Theo3}, this local minimizer is obtained from the data by a procedure that does not use the numbers of communities. In the same spirit, the algorithm in Section B.1 of the Supplementary Material starts from a preliminary embedding and from the centers derived from it, as described in Section B.3 of the Supplementary Material. The proofs of Theorem \ref{Theo3} and Corollary \ref{Cor2} are given in Sections A.4 and A.5 of the Supplementary Material.

We finally examine whether the rates and the separation requirements above can be improved.
\begin{Mth}\label{Theo4}
	There exist positive constants $c_L$ and $C_L$ depending only on $(p,q,c_0)$ such that the following holds. If $n\geqslant\max\{8(p+q)r,\,100\}$ and $s_nn^{p+q-1}\geqslant C_Lr^{\max\{2(p+q)-3,\,p+q\}}$, then every estimator $\check{\bm{\Theta}}$ of $\bm{\Theta}^*$, that is, every measurable function of $\bm{A}$, satisfies
	\begin{equation}\nonumber
		\begin{aligned}
			\sup_{\bm{\Theta}^*\in\Omega_{\Theta}}\pr\Bigl(h^2(\check{\bm{\Theta}},\bm{\Theta}^*)\geqslant\frac{c_Lr}{n^{p+q-1}}\Bigr)&\geqslant\frac{1}{4},\\
			\sup_{\bm{\Theta}^*\in\Omega_{\Theta}}\pr\Bigl(\frac{1}{\phi(n,p,q)}\|\check{\bm{\Theta}}-\bm{\Theta}^*\|_{\Omega}^2\geqslant\frac{c_Lr}{s_nn^{p+q-1}}\Bigr)&\geqslant\frac{1}{4},
		\end{aligned}
	\end{equation}
	where each probability is computed under the model with true tensor $\bm{\Theta}^*$. If, in addition, $r\geqslant 2$, then the two bounds remain valid when $\Omega_{\Theta}$ is replaced by its subset $\Omega_{\Theta}^{\mathrm{com}}$ of tensors that, together with some tail and head partitions into $r$ communities each, satisfy Condition \ref{balan} with $C_0=1/2$ and have $\Delta_n\geqslant\delta_r$, where $\delta_r=c_0^{p+q}16^{-p}r^{-(p+q-1)/2}$.
\end{Mth}
Theorem \ref{Theo4} shows that no estimator improves the rates $\frac{r}{n^{p+q-1}}$ and $\frac{r}{s_nn^{p+q-1}}$ uniformly over $\Omega_{\Theta}$, and its second statement shows that the same holds over tensors with well-separated communities. Every tensor in $\Omega_{\Theta}^{\mathrm{com}}$ has $K_{\mathrm{ta}}^*=K_{\mathrm{he}}^*=r$ and $\Delta_n\geqslant\delta_r$, so it satisfies Conditions \ref{balan} and \ref{profile_sep} for some $\lambda_3$ and $\lambda_4$ whenever $C^2\{\frac{r^2\log n}{s_nn^{p+q-1}}+\frac{r^3}{n^2}\}<\delta_r^2$. The lower bounds therefore hold over models to which Theorem \ref{Theo2} and Corollary \ref{Cor1} apply. The oracle estimator attains these rates up to the factor $\log n$ by Theorem \ref{Theo1}, and so does the maximum likelihood estimator uniformly over $\Omega_{\Theta}$, as shown in Section A.4 of the Supplementary Material. When $r$ is of the constant order, the requirement $s_nn^{p+q-1}\geqslant C_Lr^{\max\{2(p+q)-3,\,p+q\}}$ is implied by Condition \ref{profile_sep} for all sufficiently large $n$.
\begin{Mth}\label{Theo5}
	There exist positive constants $c_E$ and $\Delta_0$ depending only on $(p,q,c_0)$ such that the following holds for every $n\geqslant 64$ and every $\Delta\in(0,\Delta_0]$. There is a finite family of models with $K_{\mathrm{ta}}^*=K_{\mathrm{he}}^*=2$ that satisfy Condition \ref{balan} with $C_0=1/2$ and Condition \ref{cohesion} with $\varsigma_n=0$ and have $\Delta_n=\Delta$, with the following property. If $s_nn^{p+q-1}\Delta^2\leqslant c_E\log n$, then every estimator of the tail partition differs from the true tail partition with probability at least $1/2$ under at least one model of the family. The same holds for the head partition.
\end{Mth}
Theorem \ref{Theo5} shows that exact recovery of the partitions requires $s_nn^{p+q-1}\Delta_n^2\geqslant c_E\log n$, even when the numbers of communities are known and the within-community variation vanishes. When $\Delta_n$ is of the constant order, this condition requires the expected number of hyperedges containing a given node, which is of order $s_nn^{p+q-1}$, to grow at least logarithmically in $n$. This is the regime in which sharp thresholds for exact recovery are known for undirected hypergraph stochastic block models \citep{zhang2023exact,dumitriu2026optimal}. When $r$ and $K_{\mathrm{ta}}^*\vee K_{\mathrm{he}}^*$ are of the constant order and $s_nn^{p+q-3}=O(\log n)$, which always holds when $p+q\leqslant 3$, Condition \ref{profile_sep} requires $s_nn^{p+q-1}\Delta_n^2$ to exceed a constant multiple of $\log n$, which is the same order. In this regime, the separation under which Theorem \ref{Theo2} holds is therefore of the necessary order. To our knowledge, the methods compared in Table~\ref{tab:related} have no counterpart to Theorems \ref{Theo3}--\ref{Theo5} or Corollary \ref{Cor2}. Their guarantees for unknown numbers of communities concern one local minimizer and include no lower bounds. Theorems \ref{Theo4} and \ref{Theo5} are proved in Sections A.7 and A.8 of the Supplementary Material.
	
\section{Simulation}
All simulation experiments use $p=q=2$, $n\in\{100,200\}$, and $r=4$, and each reported quantity is computed over $100$ independent replications. In each replication, the tail and head labels are generated independently as $\psi_{\mathrm{ta}}^*(i)\sim\operatorname{Categorical}(\pi_{\mathrm{ta}})$ and $\psi_{\mathrm{he}}^*(i)\sim\operatorname{Categorical}(\pi_{\mathrm{he}})$, $i=1,\ldots,n$. Given $K_{\mathrm{ta}}^*$ and $K_{\mathrm{he}}^*$, the tail centers $\bm{\mu}_1^*,\ldots,\bm{\mu}_{K_{\mathrm{ta}}^*}^*\in\mathbb{R}^4$ and head centers $\bm{\nu}_1^*,\ldots,\bm{\nu}_{K_{\mathrm{he}}^*}^*\in\mathbb{R}^4$ are sampled from $N_4(\bm{0},\bm{I}_4)$ by rejection sampling: within each side, no two centers may have the same coordinate-wise sign pattern and every pair of centers must have Euclidean distance at least $2$. 
Conditional on labels and centers, the real-node embeddings are $\bm{\beta}_i^*=\bm{\mu}_{\psi_{\mathrm{ta}}^*(i)}^*+\bm{\varepsilon}_i^{\mathrm{ta}}$ and $\bm{\zeta}_i^*=\bm{\nu}_{\psi_{\mathrm{he}}^*(i)}^*+\bm{\varepsilon}_i^{\mathrm{he}}$, where $\bm{\varepsilon}_i^{\mathrm{ta}}\sim N_4(\bm{0},\sigma_{\mathrm{ta}}^2\bm{I}_4)$ and $\bm{\varepsilon}_i^{\mathrm{he}}\sim N_4(\bm{0},\sigma_{\mathrm{he}}^2\bm{I}_4)$. The null embeddings are fixed at $\bm{\beta}_0^*=\frac{1}{2}\bm{1}_4$ and $\bm{\zeta}_{n+1}^*=\frac{1}{2}\bm{1}_4$. For each legal canonical tuple $\omega=(i_1,i_2;j_1,j_2)\in\Omega_{2,2}^{\mathrm{ord}}$, we set $\theta_\omega^*=\sum_{k=1}^4\beta_{i_1k}^*\beta_{i_2k}^*\zeta_{j_1k}^*\zeta_{j_2k}^*$, $\varrho_\omega^*=s_n\{1+\exp(-\theta_\omega^*)\}^{-1}$, and sample $a_\omega\sim\operatorname{Bernoulli}(\varrho_\omega^*)$ independently.

The four examples all use $(K_{\mathrm{ta}}^*,K_{\mathrm{he}}^*)=(3,4)$ but vary the membership distributions, sparsity, and latent noise. Because the tail and head labels are generated independently and even have different true numbers of communities, the target is two-sided: no single partition summarizes both sides. The simulation question is therefore whether the tail/head asymmetry remains recoverable when we introduce imbalance, a sparser hypergraph through a smaller $s_n$, or lower latent noise. Example 1 is the balanced setting with $\pi_{\mathrm{ta}}=(1/3,1/3,1/3)$, $\pi_{\mathrm{he}}=(1/4,1/4,1/4,1/4)$, $s_n=0.4$, and $(\sigma_{\mathrm{ta}},\sigma_{\mathrm{he}})=(0.3,0.3)$. Example 2 keeps $s_n=0.4$ and $(\sigma_{\mathrm{ta}},\sigma_{\mathrm{he}})=(0.3,0.3)$ but introduces two-sided imbalance, $\pi_{\mathrm{ta}}=(0.5,0.3,0.2)$ and $\pi_{\mathrm{he}}=(0.3,0.3,0.2,0.2)$. Example 3 has the same balanced membership and noise levels as Example 1 but is sparser, with $s_n=0.2$. Example 4 has the same balanced membership and sparsity as Example 1 but lower latent noise, with $(\sigma_{\mathrm{ta}},\sigma_{\mathrm{he}})=(0.1,0.1)$. 

We compare FOLD with five alternatives. Decoupled-GMM first fits the directed likelihood-only embeddings and then applies Gaussian-mixture clustering with BIC separately to the fitted tail and head embeddings. Decoupled-KM uses the same first-stage likelihood-only embeddings as Decoupled-GMM, but replaces Gaussian-mixture clustering with eigengap-selected $K$-means. Projection is a two-sided spectral clustering baseline that constructs separate tail- and head-side weighted graphs from the directed hyperedges. HEM symmetrizes the directed hypergraph into an undirected $4$-uniform hypergraph and fits a fusion-regularized embedding model \citep{zhen2023community}. DiPH is another undirected hypergraph modeling method. Unlike HEM, it uses a different undirected likelihood after symmetrizing the observed directed hyperedges \citep{yu2025modeling}.

For $s\in\{\mathrm{ta},\mathrm{he}\}$, let $\widehat K_s^{(m)}$ denote the estimated number of communities in the $m$-th replication, $m=1,\ldots,100$. We report the community-number accuracy $\mathrm{Acc}_{K,s}=\frac{1}{100}\sum_{m=1}^{100}\mathbb{I}\{\widehat K_s^{(m)}=K_s^*\}$. We also report the scaled Hamming error $H_s=\min_{\tau}\frac{1}{n}\sum_{i=1}^n\mathbb{I}\{\psi^*_{s}{(i)}\neq \tau(\widehat \psi_{s}(i))\}$, where $\tau$ ranges over maps from the estimated community labels to $\{0,1,\ldots,K_s^*\}$ under which each true label has at most one preimage. A node whose estimated community is mapped to $0$ is therefore counted as an error, and the optimal $\tau$ is a permutation when $\widehat K_s=K_s^*$. In addition, we report the adjusted Rand index $\mathrm{ARI}_s$, with details in Section B.2 of the Supplementary Material, which measures how well the estimated partition matches the true partition. A larger ARI indicates better community recovery. To evaluate probability recovery, we use $\mathrm{RMSE}_p=\left\{|\Omega_{2,2}^{\mathrm{ord}}|^{-1}\sum_{\omega\in\Omega_{2,2}^{\mathrm{ord}}}(\widehat{\varrho}_\omega-\varrho_\omega^*)^2\right\}^{1/2}$, where  $\widehat{\varrho}_\omega$ is the fitted probability for tuple $\omega$. Decoupled-KM shares the same likelihood-only fitted embeddings as Decoupled-GMM, so the two methods have the same $\mathrm{RMSE}_p$. Projection is a clustering-only baseline and is not included in the probability recovery comparison.

\begin{table}[ht]
	\centering
	\tiny
		\caption{Simulation results for $n=100$ over $100$ replications. The $\mathrm{Acc}_{K,s}$ entries are means. The remaining entries are mean (standard deviation).}
	\label{tab:simulation-main}
	\begin{tabular}{llccccccc}
		\toprule
		\multirow{2}{*}{Example} & \multirow{2}{*}{Method} & \multicolumn{7}{c}{Metric} \\
		\cmidrule(lr){3-9}
        & & $\mathrm{Acc}_{K,\mathrm{ta}}$ & $\mathrm{Acc}_{K,\mathrm{he}}$ & $H_{\mathrm{ta}}$ & $H_{\mathrm{he}}$ & $\mathrm{ARI}_{\mathrm{ta}}$ & $\mathrm{ARI}_{\mathrm{he}}$ & $\mathrm{RMSE}_p$ \\
		\midrule
			\multirow{6}{*}{Example 1} & FOLD & $\mathbf{0.780}$ & $\mathbf{0.750}$ & $\mathbf{0.058(0.089)}$ & $\mathbf{0.082(0.114)}$ & $\mathbf{0.906(0.127)}$ & $\mathbf{0.863(0.164)}$ & $\mathbf{0.146(0.028)}$ \\
			& Decoupled-GMM & $0.270$ & $0.390$ & $0.174(0.133)$ & $0.119(0.106)$ & $0.783(0.155)$ & $0.833(0.143)$ & $0.150(0.027)$ \\
			& Decoupled-KM & $0.290$ & $0.020$ & $0.219(0.136)$ & $0.361(0.152)$ & $0.685(0.193)$ & $0.499(0.225)$ & $0.150(0.027)$ \\
			& Projection & $0.080$ & $0.010$ & $0.308(0.100)$ & $0.454(0.093)$ & $0.516(0.156)$ & $0.387(0.149)$ & -- \\
			& HEM & $0.020$ & $0.000$ & $0.794(0.180)$ & $0.802(0.145)$ & $0.061(0.118)$ & $0.069(0.116)$ & $0.160(0.020)$ \\
			& DiPH & $0.160$ & $0.180$ & $0.606(0.034)$ & $0.672(0.036)$ & $0.003(0.012)$ & $0.007(0.015)$ & $0.242(0.013)$ \\
		\midrule
			\multirow{6}{*}{Example 2} & FOLD & $\mathbf{0.810}$ & $\mathbf{0.700}$ & $\mathbf{0.061(0.089)}$ & $\mathbf{0.094(0.117)}$ & $\mathbf{0.884(0.158)}$ & $\mathbf{0.850(0.166)}$ & $\mathbf{0.142(0.028)}$ \\
			& Decoupled-GMM & $0.340$ & $0.510$ & $0.168(0.136)$ & $0.118(0.115)$ & $0.764(0.196)$ & $0.827(0.158)$ & $0.146(0.027)$ \\
			& Decoupled-KM & $0.280$ & $0.030$ & $0.183(0.122)$ & $0.321(0.149)$ & $0.704(0.192)$ & $0.551(0.224)$ & $0.146(0.027)$ \\
			& Projection & $0.080$ & $0.000$ & $0.261(0.133)$ & $0.405(0.089)$ & $0.556(0.223)$ & $0.431(0.138)$ & -- \\
			& HEM & $0.000$ & $0.000$ & $0.794(0.179)$ & $0.795(0.146)$ & $0.046(0.096)$ & $0.068(0.118)$ & $0.163(0.024)$ \\
			& DiPH & $0.150$ & $0.120$ & $0.513(0.060)$ & $0.651(0.034)$ & $0.017(0.032)$ & $0.007(0.016)$ & $0.243(0.013)$ \\
	    \midrule
			\multirow{6}{*}{Example 3} & FOLD & $\mathbf{0.770}$ & $\mathbf{0.740}$ & $\mathbf{0.061(0.088)}$ & $\mathbf{0.093(0.112)}$ & $\mathbf{0.902(0.126)}$ & $\mathbf{0.844(0.169)}$ & $\mathbf{0.073(0.014)}$ \\
			& Decoupled-GMM & $0.310$ & $0.390$ & $0.157(0.131)$ & $0.130(0.124)$ & $0.797(0.162)$ & $0.814(0.173)$ & $0.075(0.013)$ \\
			& Decoupled-KM & $0.270$ & $0.020$ & $0.223(0.133)$ & $0.362(0.152)$ & $0.676(0.193)$ & $0.499(0.223)$ & $0.075(0.013)$ \\
			& Projection & $0.080$ & $0.010$ & $0.312(0.106)$ & $0.454(0.093)$ & $0.508(0.164)$ & $0.385(0.150)$ & -- \\
			& HEM & $0.000$ & $0.000$ & $0.839(0.152)$ & $0.845(0.107)$ & $0.037(0.089)$ & $0.035(0.064)$ & $0.079(0.010)$ \\
			& DiPH & $0.200$ & $0.110$ & $0.600(0.029)$ & $0.674(0.031)$ & $0.005(0.012)$ & $0.006(0.013)$ & $0.121(0.006)$ \\
		\midrule
		    \multirow{6}{*}{Example 4} & FOLD & $\mathbf{0.950}$ & $\mathbf{0.880}$ & $\mathbf{0.012(0.050)}$ & $\mathbf{0.036(0.094)}$ & $\mathbf{0.982(0.073)}$ & $\mathbf{0.948(0.127)}$ & $\mathbf{0.146(0.034)}$ \\
			& Decoupled-GMM & $0.360$ & $0.240$ & $0.120(0.125)$ & $0.129(0.107)$ & $0.862(0.143)$ & $0.858(0.117)$ & $0.149(0.033)$ \\
			& Decoupled-KM & $0.380$ & $0.040$ & $0.185(0.149)$ & $0.358(0.157)$ & $0.739(0.212)$ & $0.523(0.230)$ & $0.149(0.033)$ \\
			& Projection & $0.020$ & $0.000$ & $0.313(0.091)$ & $0.433(0.074)$ & $0.534(0.152)$ & $0.445(0.121)$ & -- \\
			& HEM & $0.010$ & $0.000$ & $0.749(0.197)$ & $0.773(0.140)$ & $0.077(0.158)$ & $0.070(0.124)$ & $0.157(0.022)$ \\
			& DiPH & $0.220$ & $0.130$ & $0.601(0.035)$ & $0.670(0.033)$ & $0.007(0.019)$ & $0.008(0.019)$ & $0.240(0.014)$ \\
		\bottomrule
	\end{tabular}
\end{table}
Table \ref{tab:simulation-main} shows that FOLD gives the best community-number accuracy, the smallest Hamming error, and the largest ARI on both sides in all four examples. For example, FOLD achieves an $\mathrm{Acc}_{K,\mathrm{ta}}$ of at least 0.77 across all four examples, whereas the largest value among the competing methods is 0.38. On the head side, FOLD's $\mathrm{Acc}_{K,\mathrm{he}}$ is at least 0.70, whereas all competitors fall below 0.55.
The advantage of FOLD is most visible relative to Decoupled-GMM: the likelihood-only embeddings can still yield strong probability recovery and high ARI in some replications, but the downstream Gaussian-mixture BIC often misses the correct numbers of communities. Decoupled-KM shares the same first-stage probability fit as Decoupled-GMM but performs worse in structural recovery, showing that the downstream eigengap-selected $K$-means step is not sufficient by itself. Projection preserves the two-sided distinction but ignores directed tail-head interactions. HEM and DiPH convert the directed hypergraph into an undirected object before fitting and therefore do not model the two-sided community structure. Finally, FOLD attains the best or nearly best $\mathrm{RMSE}_p$ in all examples, so its gains in community recovery come with accurate probability estimation.

\begin{table}[ht]
	\centering
	\tiny
		\caption{Simulation results for $n=200$ over $100$ replications. The $\mathrm{Acc}_{K,s}$ entries are means. The remaining entries are mean (standard deviation).}
	\label{tab:simulation-n200}
	\begin{tabular}{llccccccc}
		\toprule
		\multirow{2}{*}{Example} & \multirow{2}{*}{Method} & \multicolumn{7}{c}{Metric} \\
		\cmidrule(lr){3-9}
        & & $\mathrm{Acc}_{K,\mathrm{ta}}$ & $\mathrm{Acc}_{K,\mathrm{he}}$ & $H_{\mathrm{ta}}$ & $H_{\mathrm{he}}$ & $\mathrm{ARI}_{\mathrm{ta}}$ & $\mathrm{ARI}_{\mathrm{he}}$ & $\mathrm{RMSE}_p$ \\
		\midrule
			\multirow{6}{*}{Example 1} & FOLD & $\mathbf{0.730}$ & $\mathbf{0.770}$ & $\mathbf{0.052(0.084)}$ & $\mathbf{0.061(0.086)}$ & $\mathbf{0.920(0.121)}$ & $\mathbf{0.906(0.115)}$ & $\mathbf{0.145(0.029)}$ \\
				& Decoupled-GMM & $0.050$ & $0.130$ & $0.308(0.143)$ & $0.193(0.116)$ & $0.663(0.152)$ & $0.795(0.119)$ & $0.152(0.027)$ \\
			& Decoupled-KM & $0.340$ & $0.050$ & $0.213(0.149)$ & $0.386(0.149)$ & $0.692(0.211)$ & $0.479(0.210)$ & $0.152(0.027)$ \\
			& Projection & $0.110$ & $0.000$ & $0.308(0.102)$ & $0.460(0.071)$ & $0.526(0.170)$ & $0.396(0.125)$ & -- \\
			& HEM & $0.050$ & $0.060$ & $0.557(0.113)$ & $0.522(0.106)$ & $0.316(0.139)$ & $0.369(0.132)$ & $0.160(0.022)$ \\
			& DiPH & $0.110$ & $0.260$ & $0.625(0.023)$ & $0.698(0.019)$ & $0.002(0.007)$ & $0.002(0.007)$ & $0.242(0.012)$ \\
		\midrule
			\multirow{6}{*}{Example 2} & FOLD & $\mathbf{0.750}$ & $\mathbf{0.720}$ & $\mathbf{0.047(0.077)}$ & $\mathbf{0.077(0.102)}$ & $\mathbf{0.928(0.124)}$ & $\mathbf{0.885(0.141)}$ & $\mathbf{0.142(0.030)}$ \\
			& Decoupled-GMM & $0.050$ & $0.130$ & $0.335(0.137)$ & $0.192(0.112)$ & $0.598(0.152)$ & $0.778(0.128)$ & $0.147(0.029)$ \\
			& Decoupled-KM & $0.350$ & $0.040$ & $0.163(0.121)$ & $0.360(0.137)$ & $0.723(0.207)$ & $0.500(0.205)$ & $0.147(0.029)$ \\
			& Projection & $0.080$ & $0.000$ & $0.250(0.103)$ & $0.399(0.093)$ & $0.568(0.195)$ & $0.448(0.141)$ & -- \\
			& HEM & $0.070$ & $0.090$ & $0.527(0.108)$ & $0.436(0.111)$ & $0.300(0.146)$ & $0.430(0.160)$ & $0.164(0.025)$ \\
			& DiPH & $0.100$ & $0.110$ & $0.521(0.045)$ & $0.664(0.026)$ & $0.013(0.022)$ & $0.005(0.008)$ & $0.244(0.013)$ \\
	    \midrule
			\multirow{6}{*}{Example 3} & FOLD & $\mathbf{0.750}$ & $\mathbf{0.790}$ & $\mathbf{0.049(0.079)}$ & $\mathbf{0.060(0.085)}$ & $\mathbf{0.922(0.118)}$ & $\mathbf{0.907(0.113)}$ & $\mathbf{0.073(0.014)}$ \\
			& Decoupled-GMM & $0.060$ & $0.090$ & $0.306(0.146)$ & $0.216(0.113)$ & $0.666(0.154)$ & $0.775(0.115)$ & $0.077(0.013)$ \\
			& Decoupled-KM & $0.340$ & $0.050$ & $0.214(0.149)$ & $0.387(0.145)$ & $0.690(0.212)$ & $0.477(0.204)$ & $0.077(0.013)$ \\
			& Projection & $0.100$ & $0.000$ & $0.309(0.102)$ & $0.460(0.072)$ & $0.524(0.171)$ & $0.397(0.125)$ & -- \\
			& HEM & $0.080$ & $0.090$ & $0.532(0.127)$ & $0.526(0.108)$ & $0.308(0.159)$ & $0.329(0.163)$ & $0.080(0.012)$ \\
			& DiPH & $0.140$ & $0.230$ & $0.623(0.024)$ & $0.697(0.024)$ & $0.003(0.009)$ & $0.002(0.007)$ & $0.121(0.006)$ \\
		\midrule
		    \multirow{6}{*}{Example 4} & FOLD & $\mathbf{0.940}$ & $\mathbf{0.930}$ & $\mathbf{0.015(0.064)}$ & $\mathbf{0.014(0.057)}$ & $\mathbf{0.979(0.092)}$ & $\mathbf{0.983(0.067)}$ & $\mathbf{0.148(0.036)}$ \\
			& Decoupled-GMM & $0.030$ & $0.070$ & $0.272(0.135)$ & $0.214(0.113)$ & $0.708(0.142)$ & $0.790(0.113)$ & $0.150(0.034)$ \\
			& Decoupled-KM & $0.470$ & $0.030$ & $0.165(0.157)$ & $0.392(0.131)$ & $0.771(0.218)$ & $0.484(0.193)$ & $0.150(0.034)$ \\
			& Projection & $0.090$ & $0.000$ & $0.306(0.117)$ & $0.451(0.062)$ & $0.568(0.182)$ & $0.444(0.107)$ & -- \\
			& HEM & $0.010$ & $0.080$ & $0.593(0.115)$ & $0.553(0.090)$ & $0.329(0.112)$ & $0.390(0.110)$ & $0.162(0.023)$ \\
			& DiPH & $0.170$ & $0.180$ & $0.624(0.020)$ & $0.695(0.023)$ & $0.002(0.007)$ & $0.003(0.007)$ & $0.240(0.013)$ \\
		\bottomrule
	\end{tabular}
\end{table}
Table \ref{tab:simulation-n200} gives the corresponding results when the network size is increased to $n=200$. The main pattern remains the same: FOLD has the highest community-number accuracy, the smallest Hamming error, and the largest ARI on both the tail and head sides in all four examples. Compared with Table \ref{tab:simulation-main}, the improvement is especially clear for the head-side partition recovery, where the Hamming errors of FOLD decrease substantially in Examples 1, 3, and 4. The likelihood-only baselines still provide competitive probability recovery, but the community-number accuracy of Decoupled-GMM decreases as the sample size increases. Projection and Decoupled-KM are useful two-sided clustering baselines, but their eigengap-based cluster selection is not sufficiently stable. The two undirected hypergraph baselines, HEM and DiPH, remain clearly disadvantaged in recovering the directed two-sided community structure. Overall, the advantage of FOLD persists at $n=200$ and becomes more pronounced in several structural metrics as the hypergraph grows.

\section{The CSCN analysis}\label{sec:real-data}

We now turn to the  Cross-Statistics Citation Network (CSCN)  dataset to examine how statistical methodology moves between Statistics and neighboring fields, and which groups of authors send and receive it. The object of interest is not only whether two authors are close on a research map, but also whether an author plays distinct roles on the citee (tail) side versus the citer (head) side. A directed hypergraph suits this purpose because it preserves the author groups of each paper citation.

To support this analysis, we constructed the CSCN dataset, a statistics-anchored citation corpus covering papers published from 2005 through 2024. The corpus includes 16 Statistics journals and 26 venues from three neighboring fields: 10 Biology journals, 7 Computer Science venues (4 conferences and 3 journals), and 9 Economics journals. A non-Statistics paper is retained only when it has a direct citation link with the Statistics core: it either cites a Statistics paper or is cited by a Statistics paper. The full raw corpus contains 44,004 papers, 27,663 authors, and 252,912 internal reference links. To our knowledge, CSCN is the largest corpus constructed for studying citation dynamics between Statistics and its neighboring fields. The initial discovery layer is OpenAlex, but all records were repaired and cross-validated using Crossref, Semantic Scholar, PubMed, Scopus, Web of Science API Expanded, MADStat \citep{ji2022cocitation}, and official proceedings pages for major Computer Science conferences. Full details on venue scope, data cleaning, and author disambiguation are given in Sections C.1 and C.2 of the Supplementary Material. 

For the FOLD illustration, we work with a 233-author subgraph constructed to retain relatively high directed-hyperedge density while maintaining cross-field balance. Details of the subgraph construction are given in Section C.3 of the Supplementary Material. After self-overlap removal and truncation to $p=q=2$, the 8,691 paper citations among these authors expand into 107,551 directed author hyperedges. Different paper citations can generate the same canonical tuple, so these hyperedges involve 42,833 distinct tuples. Repeated tuples are kept, so that a tuple supported by several paper citations carries proportionally more weight in the fit. The density of the hyperedges is $107551/\phi(233, 2, 2)=1.47\times 10^{-4}$. Because $\phi(233,2,2)$ exceeds $7\times 10^{8}$, the likelihood is evaluated on all observed hyperedges together with ten randomly drawn unobserved tuples per hyperedge, as detailed in Section C.3 of the Supplementary Material.

\subsection{Community detection results}\label{sec:real-data-fold}

\begin{figure}[!t]
    \centering
    \subfigure[Tail-side embedding.]{
    \includegraphics[width=0.47\textwidth]{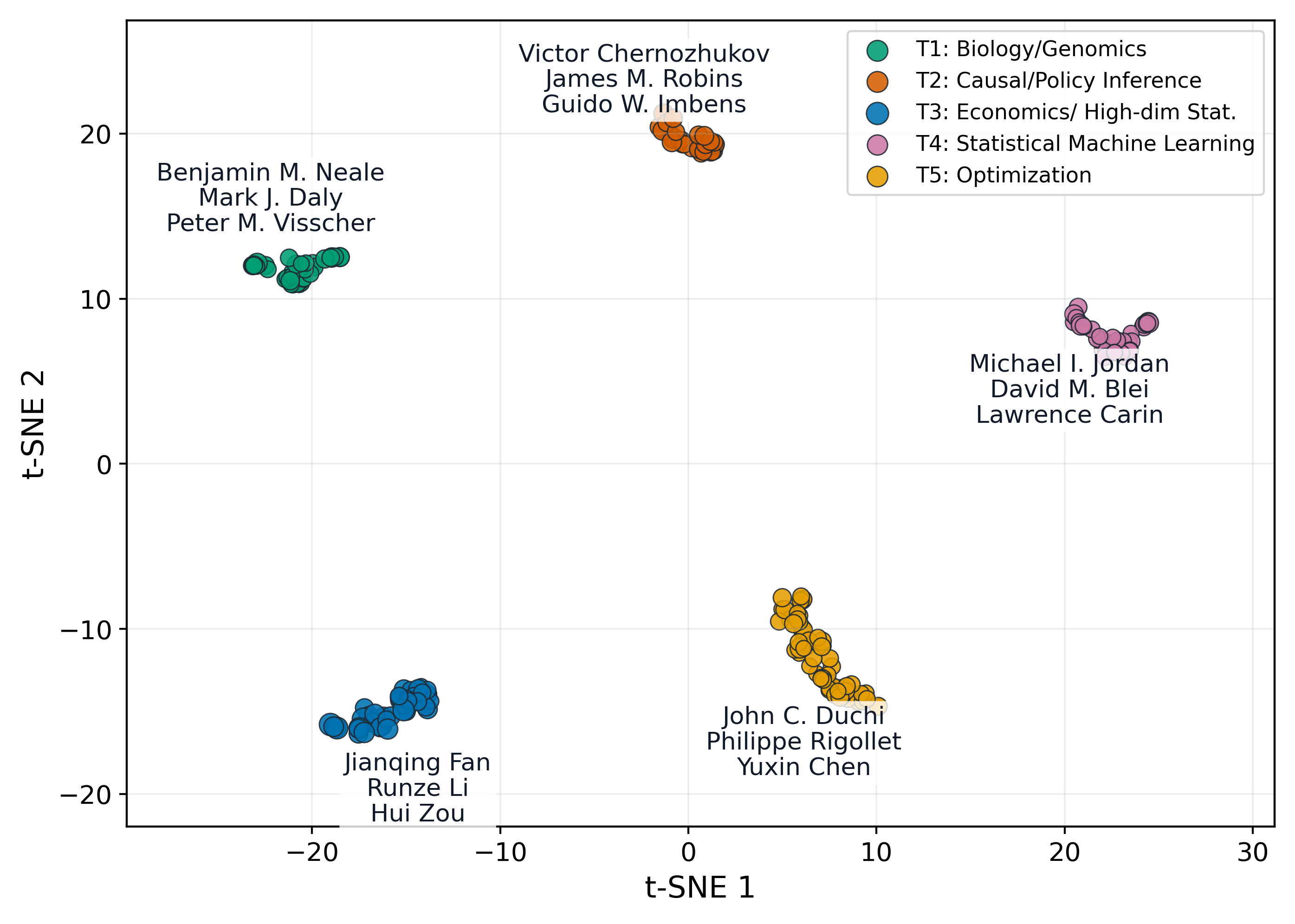}}
    \subfigure[Head-side embedding.]{
    \includegraphics[width=0.47\textwidth]{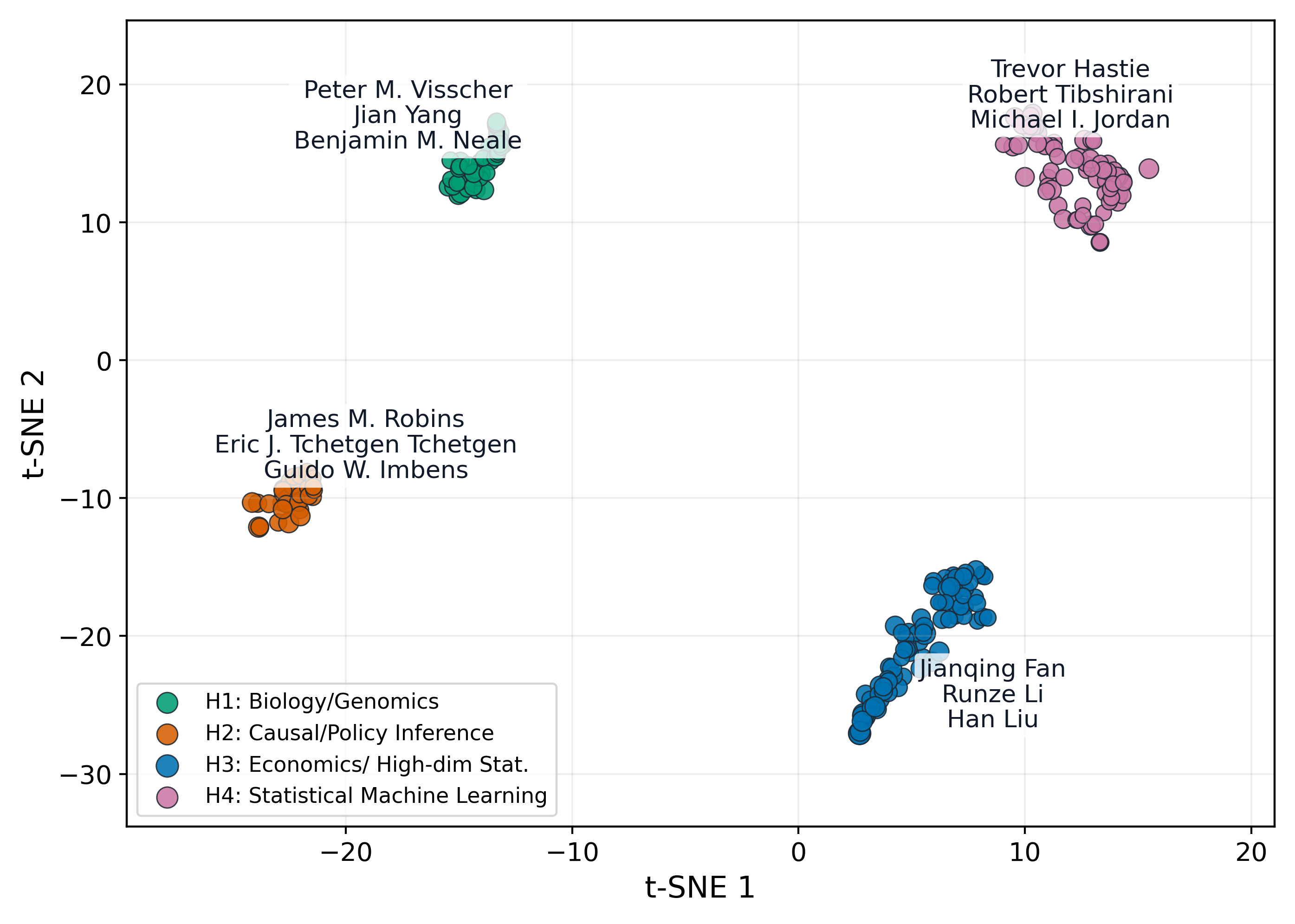}}
    \caption{FOLD embeddings for the 233-author directed hypergraph.}
    \label{fig:real-data-tsne}
\end{figure}

Applying FOLD with embedding dimension $r=4$ gives five tail and four head communities.  Figure~\ref{fig:real-data-tsne} shows the two sets of fitted embeddings after t-SNE visualization.  We assign community names after estimation by extracting keywords from paper titles, grouping them into several theme concepts, and ranking these concepts by their weighted frequencies within each fitted tail or head community. The keyword dictionary, ranking rule, and author membership table are given in Section C.4 of the Supplementary Material.

The tail communities are labeled as T1 Biology/Genomics, T2 Causal/Policy Inference, T3 Economics/High-dimensional Statistics, T4 Statistical Machine Learning, and T5 Optimization.  The corresponding head communities are H1 Biology/Genomics, H2 Causal/Policy Inference, H3 Economics/High-dimensional Statistics, and H4 Statistical Machine Learning.  These labels should be read as role labels, not as rigid author labels.  For example, a researcher may appear in a high-dimensional statistics tail role but in a statistical-machine-learning or biological-data head role.  In the fitted result, 84 of the 233 authors (36.1\%) have different tail and head roles, which supports modeling the two sides separately. 

\begin{figure}[!htbp]
\centering
\includegraphics[width=0.7\textwidth]{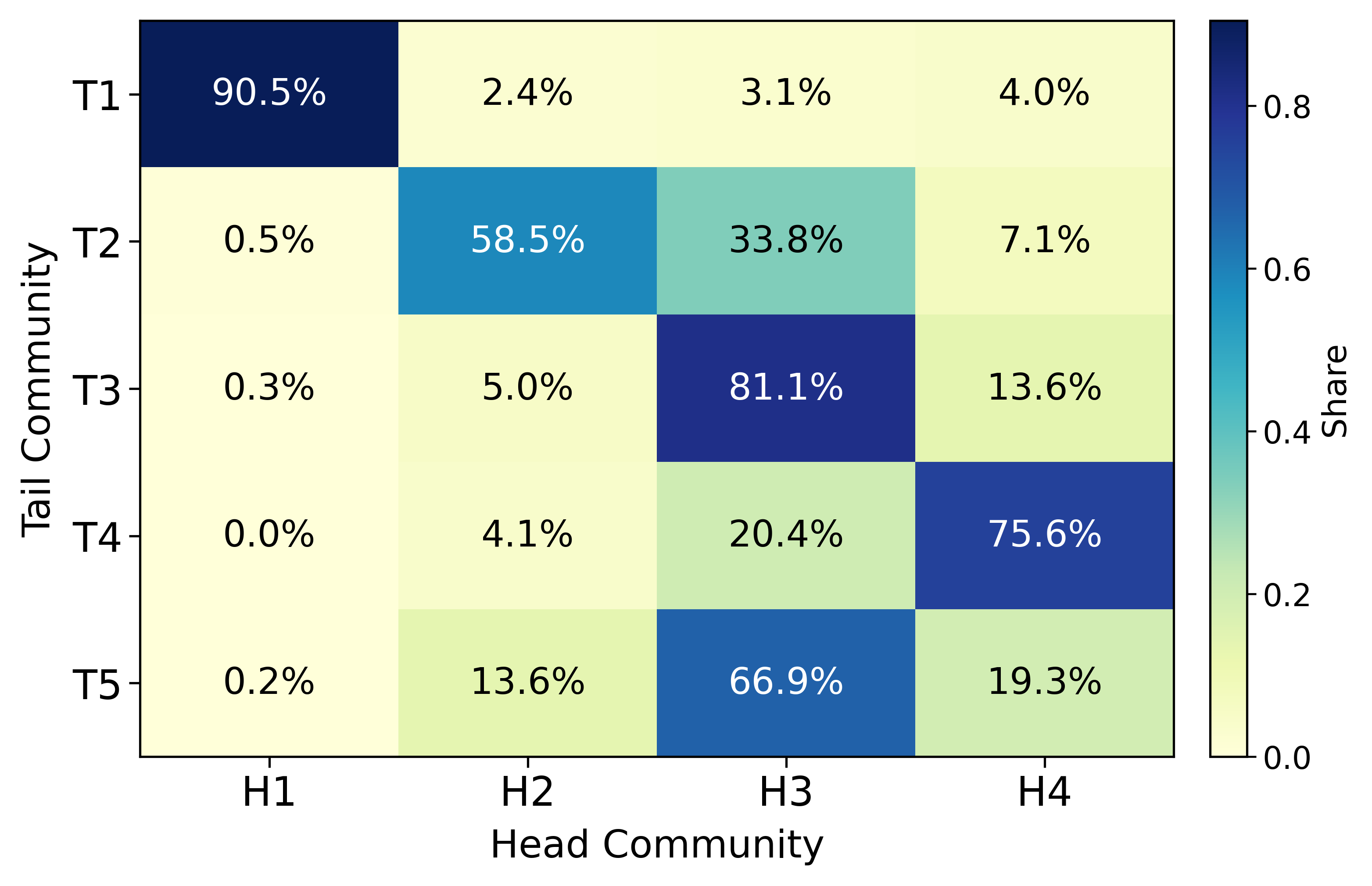}
\caption{Paper-level directional flow from tail communities (rows) to head communities (columns). Each paper citation has total weight one, split equally over the pairs of distinct tail and head communities on its two sides, and each row is normalized by its total.}
\label{fig:real-data-flow}
\end{figure}

The absence of an H5 community is informative.  In Figure~\ref{fig:real-data-flow}, T5 sends $66.9\%$ of its paper-level knowledge flow to H3, $19.3\%$ to H4, and $13.6\%$ to H2, with essentially no flow to H1.  Optimization therefore appears as a coherent tail community, but on the head side its flow is absorbed mainly by the Economics/High-dimensional Statistics and Statistical Machine Learning communities. 
A concrete example is John C. Duchi: his work on distributed and stochastic optimization, including communication-efficient algorithms, provides computational tools used to fit statistical learning models at scale \citep{zhang2013communication}, rather than defining a separate optimization receiving community.
This is precisely the asymmetry that the directed hypergraph is designed to capture: an author group can form a distinct source-side community without generating a separate receiving-side block.

Figure~\ref{fig:real-data-flow} summarizes the resulting traffic between the two maps.  The diagonal cells remain dominant, with T1$\to$H1 = $90.5\%$, T2$\to$H2 = $58.5\%$, T3$\to$H3 = $81.1\%$, and T4$\to$H4 = $75.6\%$.  The off-diagonal flows are nevertheless substantive: T2 sends $33.8\%$ of its output to H3, T3 sends $13.6\%$ to H4, and T5 sends $66.9\%$ to H3 plus $19.3\%$ to H4.  These off-diagonal shares show that the directed communities record systematic cross-role borrowing between tail and head roles.

\subsection{Scientific findings}\label{sec:real-data-findings}

\begin{figure}[!htbp]
    \centering
    \subfigure[Venue presence versus uptake.]{
    \includegraphics[width=0.54\textwidth]{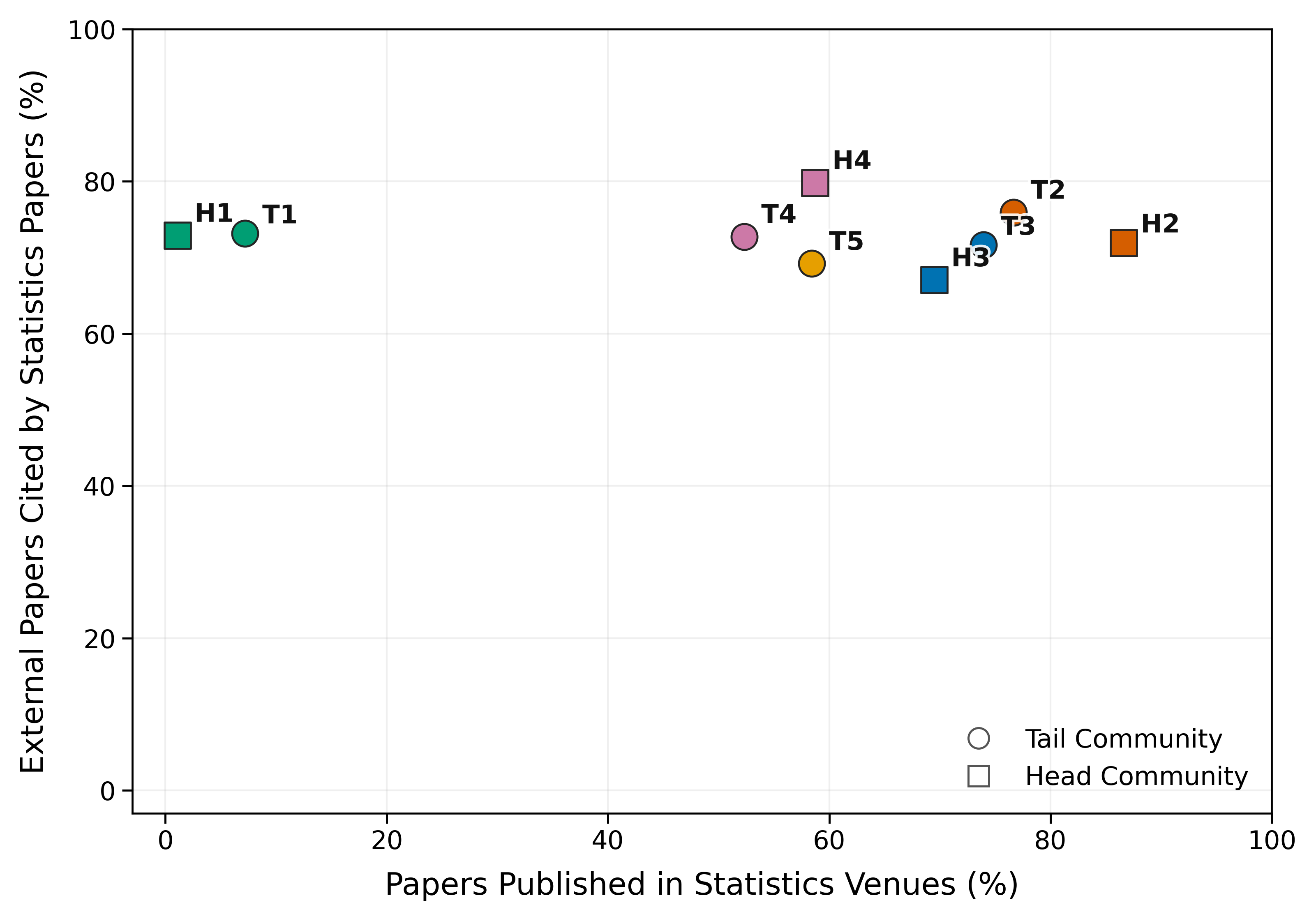}}
    \hspace{0.005\textwidth}
    \subfigure[Directional split for T1/H1.]{
    \includegraphics[width=0.30\textwidth]{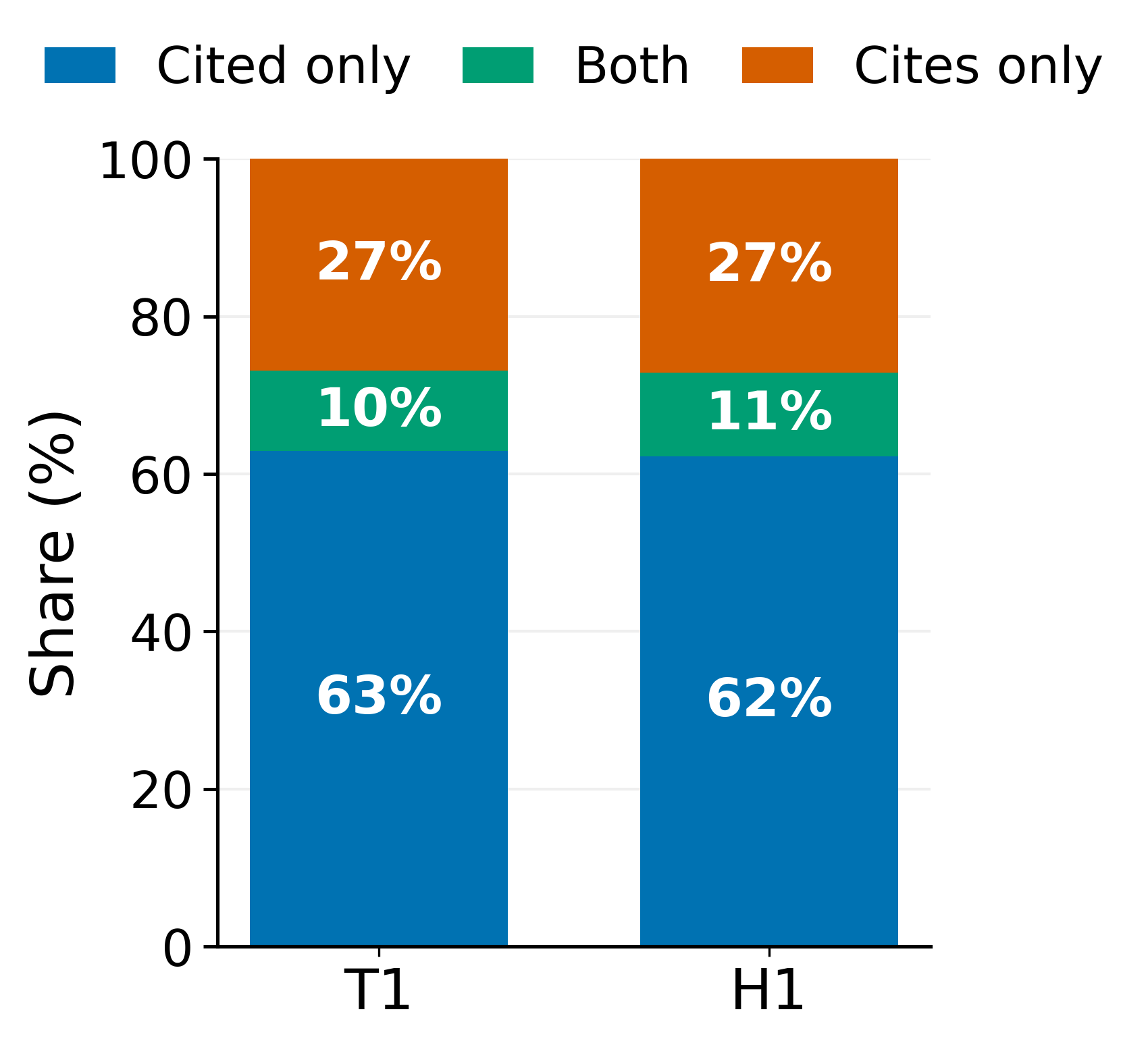}}
    \caption{Statistics linkage of FOLD communities.}
    \label{fig:real-data-stat-linkage}
\end{figure}

We next examine what the directed communities reveal about knowledge flow in the Statistics-centered corpus. We summarize the main patterns in four points.
The first point is that venue composition alone is misleading. Biology/Genomics is a clearly tail-dominant community rather than a marginal outside field.  Figure~\ref{fig:real-data-stat-linkage}(a) shows that the Biology/Genomics communities T1 and H1 publish almost no papers in Statistics venues, $7.2\%$ and $1.1\%$, respectively, yet $73.1\%$ and $72.9\%$ of their external papers are cited by Statistics.  Figure~\ref{fig:real-data-stat-linkage}(b) makes the asymmetry explicit: for T1, $62.9\%$ of non-Statistics papers are only cited by Statistics, compared with $26.9\%$ that only cite Statistics. For H1, the corresponding split is $62.3\%$ versus $27.1\%$.  These patterns show that Biology/Genomics supplies empirical systems and data objects on which statistics-core methodology is tested, calibrated, and interpreted, while the reverse direction is narrower and more tool-specific.

\begin{figure}[!htbp]
\centering
\includegraphics[width=0.80\textwidth]{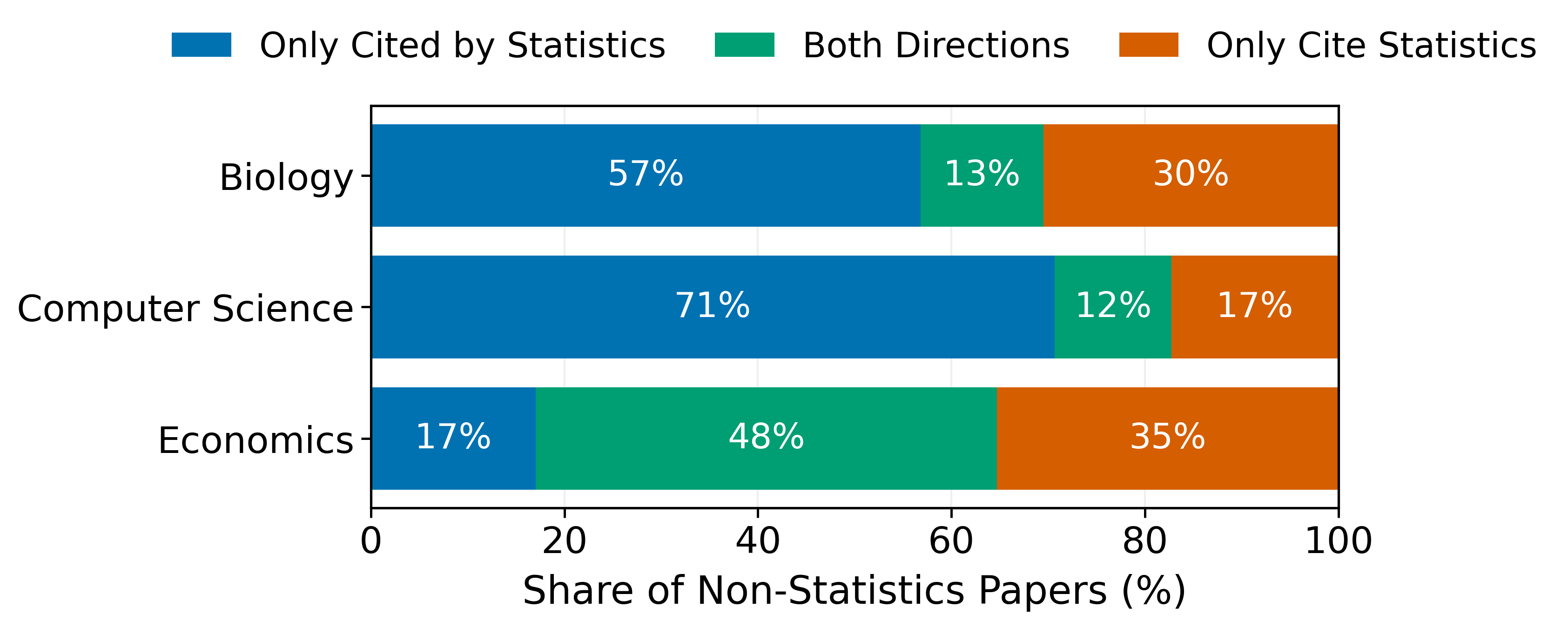}
\caption{Directional Statistics linkage of Biology, Computer Science, and Economics papers authored by the selected subgraph authors.}
\label{fig:real-data-field-direction}
\end{figure}

A second point concerns how the neighboring domains connect to Statistics in different directions. The three neighboring domains are directionally heterogeneous, and Economics is noticeably more reciprocal than Biology or Computer Science. Figure~\ref{fig:real-data-field-direction} shows that Biology and Computer Science both lean strongly toward the ``only cited by Statistics'' side, but with different magnitudes: Biology is $56.8\%$ only cited, $30.5\%$ only cites, and $12.7\%$ bidirectional, whereas Computer Science is $70.7\%$ only cited, $17.3\%$ only cites, and $12.1\%$ bidirectional. Economics behaves differently: its papers are $47.6\%$ bidirectional, $35.3\%$ only cite Statistics, and only $17.1\%$ are only cited by Statistics. The domain-level picture therefore separates tail-dominant Biology, and especially Computer Science, from the much more reciprocal exchange with Economics.

A third point is that these cross-field links evolve over time. Several off-diagonal links are clearly dynamic and reflect recent directional shifts rather than static background similarity. 
\begin{figure}[h]
    \centering
    \subfigure[Causal flow evolution.]{
    \includegraphics[width=0.47\textwidth]{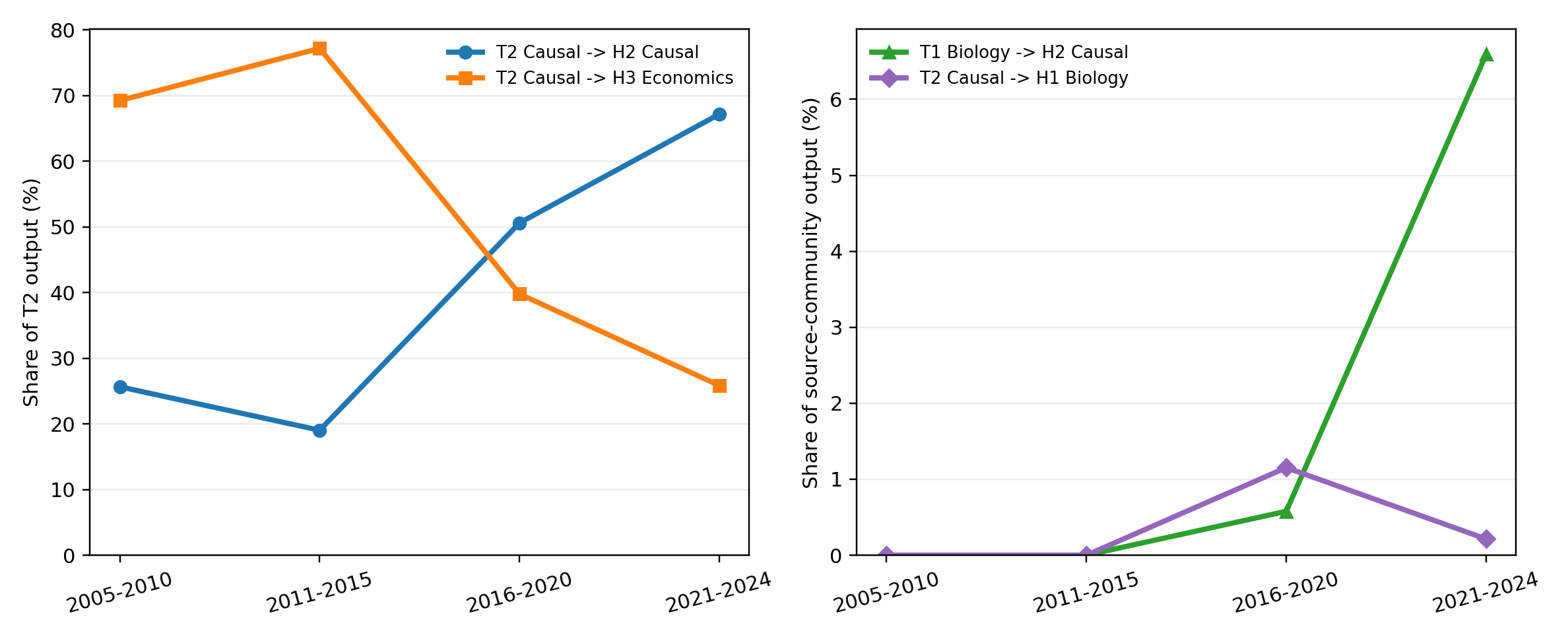}}
    \subfigure[Changing direction.]{
    \includegraphics[width=0.47\textwidth]{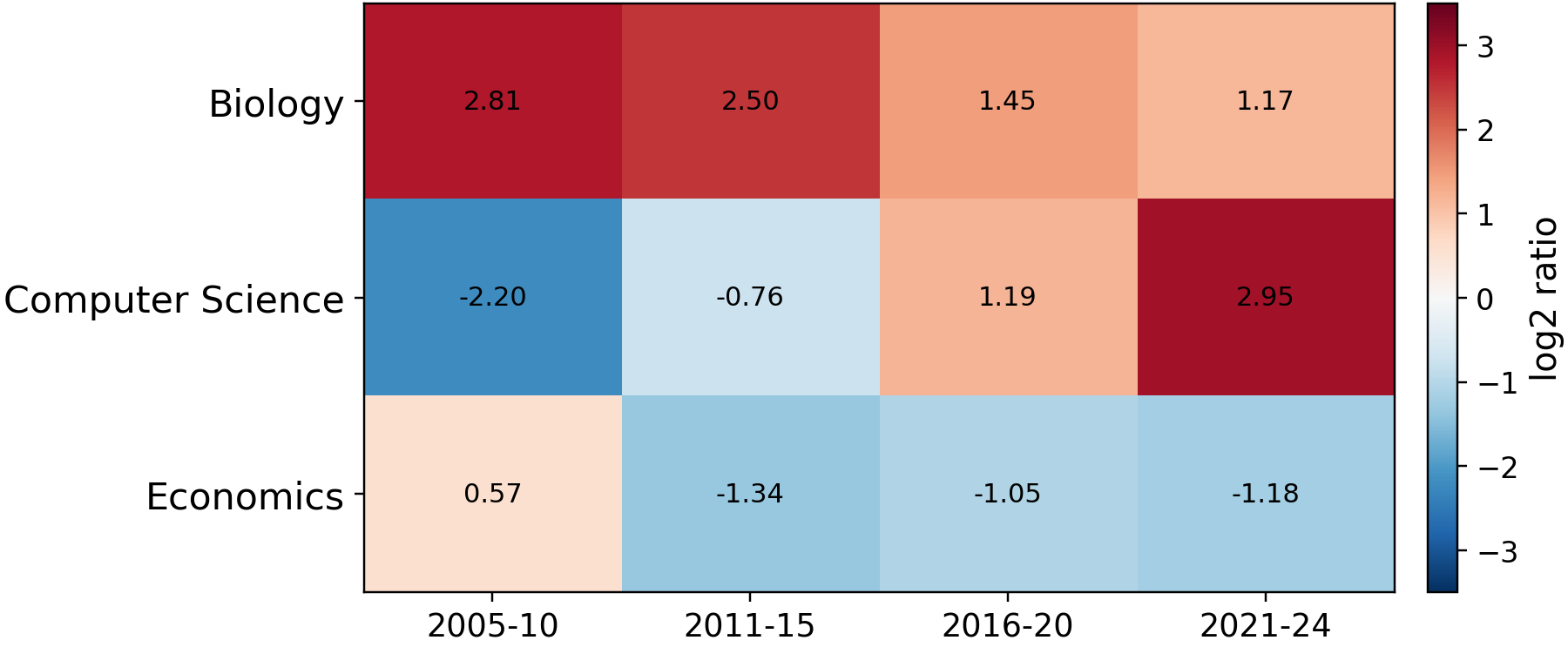}}
    \caption{Temporal changes in selected cross-field flows. (a) Share of the outgoing paper-level flow of each tail community received by each head community. (b) $\log_2\{(a+0.5)/(b+0.5)\}$, where $a$ is the number of citations from Statistics papers to papers of the field and $b$ is the number in the reverse direction.}
    \label{fig:real-data-temporal}
\end{figure}
In the left panel of Figure~\ref{fig:real-data-temporal}, the T2$\to$H3 share falls from $69.2\%$ in 2005--2010 to $25.8\%$ in 2021--2024, while T2$\to$H2 rises from $25.6\%$ to $67.2\%$. Over the same periods, the share of the T1 outflow received by H2 increases from $0\%$ to $6.6\%$, so Causal/Policy Inference papers increasingly cite Biology/Genomics work. Causal methodology, once taken up mainly in economics settings, is thus increasingly used within the causal-inference community, which in turn draws more on biological and genomic studies. The right panel shows a different temporal shift for Computer Science. In 2011--2015, Statistics papers made 31 citations to Computer Science papers, compared with 53 citations in the reverse direction. By 2021--2024, the counts were 181 and 23, a reversal in the net direction of citation flow.

\begin{figure}[!htbp]
    \centering
    \includegraphics[width=0.75\textwidth]{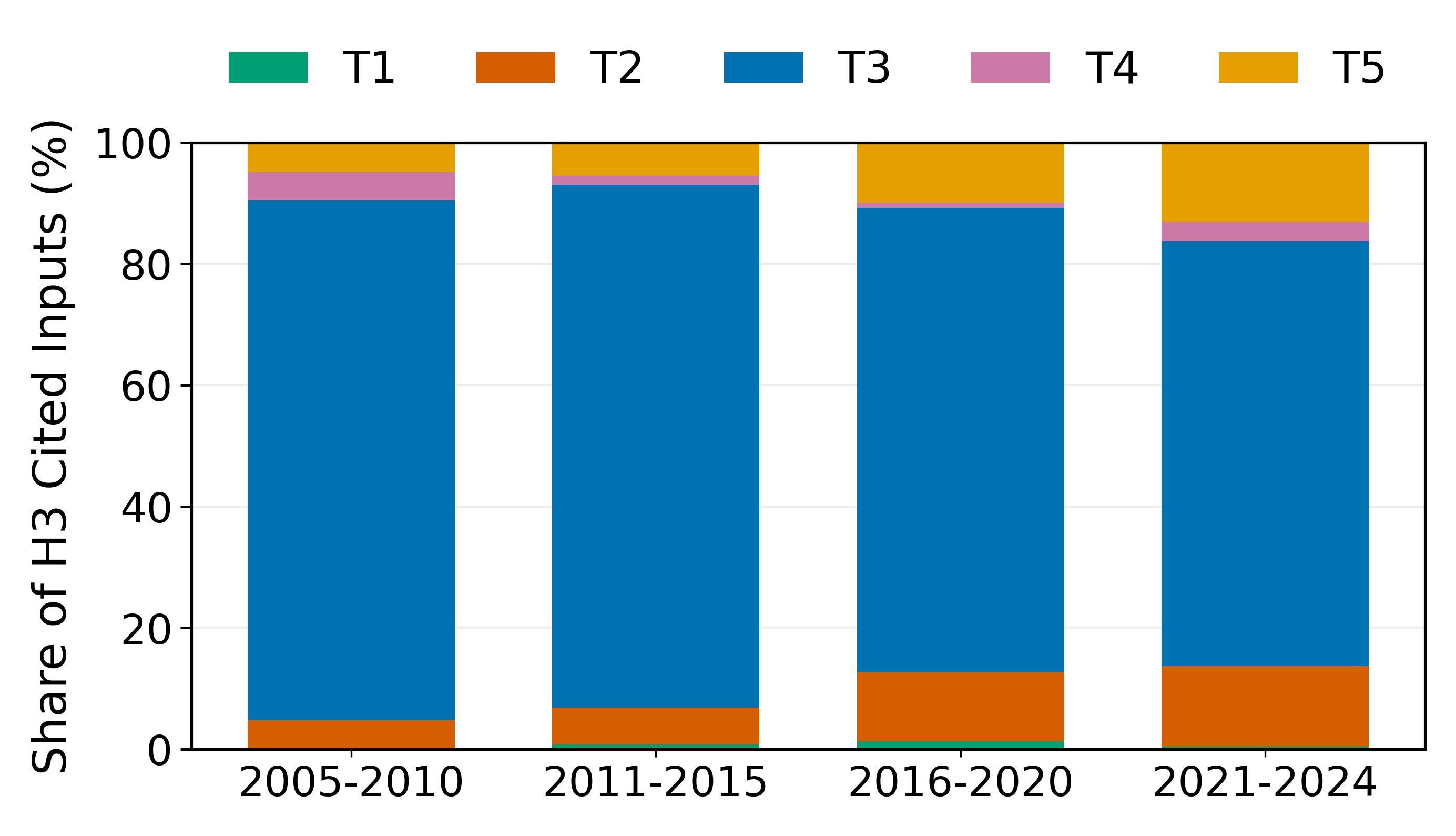}
    \caption{Changing tail-community composition of the Economics/High-dimensional Statistics head community H3 across time periods.}
    \label{fig:real-data-convergence}
\end{figure}

A fourth point is that the Economics/High-dimensional Statistics head community H3 is becoming a point of convergence for several tail communities. H3 now looks increasingly hybrid rather than like a narrow econometrics-only head block. Figure~\ref{fig:real-data-convergence} shows that, among the cited inputs received by H3, T3 remains the largest component, but its share declines from $85.7\%$ in 2005--2010 to $70.0\%$ in 2021--2024. Over the same periods, T2 rises from $4.7\%$ to $13.2\%$ and T5 rises from $4.9\%$ to $13.2\%$. H3 should therefore be read as a broadened head block that increasingly combines high-dimensional statistics with causal methodology and optimization infrastructure.

\subsection{Comparison with ordinary directed network analysis}\label{sec:real-data-alternative}

To assess the value of retaining each citation as a directed hyperedge, we compare the proposed directed hypergraph analysis with an ordinary directed network analysis on the same 233 authors, in which the projected ordinary directed network is obtained by replacing each directed hyperedge with ordinary directed edges between all author pairs on its two sides.  For example, the directed hyperedge $\{i,i'\}\to\{j,j'\}$ is replaced by the four edges $i\to j$, $i\to j'$, $i'\to j$, and $i'\to j'$.  We then analyze this graph using a network embedding model for ordinary directed graphs, with the same embedding dimension and fusion procedure as in the hypergraph analysis.  This projected ordinary directed network retains the citation direction, but it no longer records that $i$ and $i'$ appear together in one cited paper and that $j$ and $j'$ appear together in one citing paper.  The comparison shows whether keeping these author groups changes the detected communities.

To evaluate the resulting tail and head partitions, we use a pair of side-specific modularities motivated by two existing constructions.  \citet{leicht2008community} defined directed-graph modularity using the sending and receiving degrees in its null model, while \citet{kumar2018hypergraph} defined hypergraph modularity through a degree-preserving reduction.  We retain the distinction between the tail and head roles and apply the latter reduction separately to the cited and citing author sets (Section C.5 of the Supplementary Material).

The hypergraph analysis identifies five tail communities and four head communities, with corresponding modularities of $0.5776$ and $0.6335$. By comparison, the ordinary directed network analysis identifies four tail communities and three head communities, with lower modularities of $0.4981$ and $0.4955$, respectively. The finer partition produced by the hypergraph method is thus accompanied by stronger within-community concentration in the original paper-level author sets. As an additional calibration, we construct an oracle configuration that preserves the observed total degree of each estimated community while assigning all reduced edge weights within communities. The resulting Oracle modularities are $0.6971$ for the tail side and $0.6758$ for the head side. The modularities obtained from the hypergraph analysis therefore reach $82.9\%$ and $93.7\%$ of their respective Oracle benchmarks, with the head partition lying particularly close to its idealized configuration.

Compared with the hypergraph analysis, the ordinary directed network analysis removes the paper-level grouping that distinguishes repeated dyadic links from joint participation in one citing or cited author set.  In the resulting graph fit, Optimization no longer survives as a separate tail role, and the causal, economic, and machine-learning receiving patterns are compressed into fewer head communities.  By retaining each citation as a set-to-set event, the hypergraph analysis recovers finer role-specific structure while also achieving stronger within-community concentration under the same original paper-level evidence.  The calculation of the modularities, their Oracle counterparts, and the uncertainty assessment are provided in Section C.5 of the Supplementary Material.

\section{Discussion}

Our model treats each ordered many-to-many event as the basic unit of analysis, preserving tail-head roles and varying hyperedge cardinalities without dyadic projection. FOLD jointly estimates hyperedge probabilities, tail and head memberships, and their unknown and potentially unequal numbers of communities. The theory establishes recovery in sparse regimes together with lower bounds on what any method can achieve, and the simulations demonstrate stable tail and head recovery under community imbalance, sparser observations, and lower latent noise.  The newly constructed Cross-Statistics Citation Network then makes it possible to study how statistical knowledge moves across fields while retaining citation direction and paper-level author sets.  

\section*{Funding}
This work was partially supported by the National Natural Science Foundation of China (12501373), Shenzhen Medical Research Fund (E250200620, E250200621), and the Shanghai Sailing Program (24YF2721900).

\noindent{\bf Disclosure statement.} The authors declare no conflicts of interest.

\noindent{\bf Data Availability Statement.} The working network dataset used in the real-data analysis is available at the following URL: \url{https://github.com/zgj19stat/FOLD}.

\noindent{\bf Supplementary Material.} The Supplementary Material contains the proofs of all theoretical results, the computational details of the algorithm, additional simulation information, and supplementary real-data analyses.

\bibliography{reference}

\end{document}